\documentclass[aps,prl,twocolumn,superscriptaddress,groupedaddress,amsmath,amssymb,nofootinbib]{revtex4-2}
\usepackage{graphicx}%
\usepackage{bm}        %
\usepackage{amssymb}
\usepackage[normalem]{ulem}
\usepackage{booktabs}  %
\usepackage{makecell}  %
\usepackage{tikz}      %
\usetikzlibrary{arrows.meta,positioning}
\usepackage[usenames,dvipsnames]{xcolor}
\usepackage{color}
\usepackage{url}
\RequirePackage[hidelinks]{hyperref}

\def\beq{\begin{equation}}
\def\eeq{\end{equation}}

\newcommand{\Qout}{Q_{\mathrm{out}}}

\begin{document}

\title{An anatomically grounded human-scale steady hydraulic model of the lymphatic network}

\author{Chia-Feng Chang}
\author{Ayemon Baraka}
\author{Egor Sidorov}
\author{Sabee Molloi}
\email{symolloi@uci.edu}
\affiliation{Department of Radiological Sciences, University of California, Irvine, CA, USA}
\date{\today}

\begin{abstract}
\noindent\textbf{Background} The lymphatic system returns $2$--$3$\,L of interstitial fluid per day to the venous circulation and drains the central nervous system. Models of lymph transport are almost all reduced-order, one lymphangion or a short chain, so they cannot show how body-wide topology, calibre hierarchy and the venous outlet set the pressure and flow field. No whole-body, anatomically explicit description exists.

\noindent\textbf{Methods} We build a passive hydraulic model on an anatomically derived graph of the human lymphatic system, 1010 vertices and 1126 edges expanded to 89{,}060 sub-edges. Radii are closed by a Murray cube law with two independent derivations, minimum work and endothelial wall-shear-stress homeostasis. Pressure and flow follow from nodal hydraulic balance over Hagen--Poiseuille conductances, an exponential wall law, hysteretic valves, Starling leak, node retention and a Windkessel outlet. Active contraction is off, isolating the hydraulic skeleton.

\noindent\textbf{Results} The model resolves steady transmural pressure, flow and radius on every sub-edge. Two independent quantities land in clinical range at the same operating point, and neither is fitted. Radii from the Murray closure fall inside the vessel-class range, from the $1.50$\,mm thoracic duct to the $0.16$\,mm cervical collectors. Pressures from the nodal solve stay below the $50$\,mmHg cuff scale on all sub-edges. About $49$\% of the gross load is removed before the outlet, nearly independently of the outflow scale and of the outlet parameters. On the route draining the central nervous system, $30$\% of the superficial cervical load reaches the vein, at $10$ to $18$\,mmHg along a segment cervical-scale models replace with a prescribed outlet pressure.

\noindent\textbf{Conclusions} This work moves the unit of lymphatic modelling from the vessel to the body. One nodal Kirchhoff system over the anatomical territories makes pressure and flow on every segment follow from the whole network, not from a prescribed boundary condition, with only the inflow scale calibrated. It is therefore a physiologically consistent passive baseline built from literature-grounded physics, and a substrate for contraction-driven and clinical extensions.%

\end{abstract}

\maketitle


\section{Background}

The lymphatic system returns interstitial fluid, protein, lipid, and immune cells to the venous circulation, draining $\sim$2--3\,L of lymph per day into the subclavian veins through the thoracic duct \cite{GuytonHall2015,Moazzam2026}. Unlike the blood circulation, lymph transport is driven both by extrinsic tissue forces and by the intrinsic, valve-gated contraction of muscular vessel segments (lymphangions) \cite{Zawieja2009,davis2025control,Davis2009}. Quantitative models of this transport have been pivotal in interpreting experiments, but the modelling literature is overwhelmingly \emph{reduced-order}: single lymphangions or short chains of a few lymphangions with lumped parameters \cite{Bertram2014,Davis2011,MargarisBlack2012,Razavi2017,jamalian2016network,ikhimwin2020computational,KimTithof2024}. Such models capture single-vessel pump physics but cannot represent how body-wide topology, hierarchical vessel calibre, and a realistic venous outlet jointly determine the pressure and flow distribution. The same network also drains the central nervous system. Cervical lymphatic function declines with age, and cerebrospinal-fluid drainage declines with it \cite{du2024restoration}. After traumatic brain injury a drop in that contractility, together with a rise in central venous pressure, drives cerebral oedema \cite{Hussain2023}.

The modelling landscape has recently been surveyed across the initial-, collecting-, and node-scale levels, with whole-body integrated network models identified as a principal open direction \cite{JayathungageDon2024}. One whole-body graph model of the human lymphatic system already exists. It converts an anatomical graph into an oriented one by solving a steady Poiseuille and mass-conservation balance at the vessel junctions \cite{Savinkov2020}, so the topology and the direction of flow are already fixed there. Its reported outputs are graph metrics such as energy, clustering and robustness, not physical radii, valves, wall mechanics or a pressure field. In parallel, recent experiments continue to sharpen the constitutive inputs such models require, including direct pressure--flow and valve-resistance measurements on the thoracic duct itself \cite{PatelLuKassab2025}, servo-null pressures across mouse lymphatic networks \cite{Davis2026_MouseLymphaticPressure}, and resting human thoracic-duct calibre by ultrasound \cite{Shinaoka2025}.

Extending to a whole-body, anatomically explicit network raises a problem that is trivial in reduced-order work but acute at scale: how to assign a radius to each of the thousands of vessel segments. Fitting every edge is both underdetermined and physiologically uninformative. We instead close the radius assignment with \emph{Murray's law}, the cube relationship between flow and calibre \cite{Murray1926}. Murray's law is usually invoked as an empirical branching rule. We show that for the lymphatic network it is supported by \emph{two independent physical derivations}: (1) a minimum-work optimality argument and (2) an endothelial wall-shear-stress homeostasis argument, which converge on the same cube exponent and whose consistency condition. This dual grounding is, to our knowledge, not previously stated for a lymphatic application and is a central methodological contribution.

In this work we deliberately restrict attention to the \emph{static}, passive regime: active circumferential muscle tension is switched off. This isolates the hydraulic skeleton of the model which includes network topology, Murray closure, passive wall constitutive behaviour, valve resistance, transmural leak, lymph-node retention, and the venous-outlet boundary condition from the contractile machinery, whose calibration is a separate and more delicate problem reserved for a companion study. We emphasis that the overall inflow scale is calibrated so that the net thoracic-duct outflow matches the physiological observation. The genuinely untuned, falsifiable outputs of the static model are instead the spatial pressure distribution and the per-region calibre hierarchy, which can be tested against clinical data. \cite{Scallan2016,Breslin2019,MooreBertram2018,Solari2020,Thorup2023}

The specific contributions of this paper are: (i) a human-scale, anatomically derived lymphatic network model with an explicit three-level discretization (edge, lymphangion, sub-segment); (ii) a Murray radius closure justified by two independent derivations with a stated consistency condition, and its per-branch flow-weighted form; (iii) a passive steady solver coupling Hagen-Poiseuille conductances, an exponential wall law, hysteretic valves, Starling leak, and lymph-node retention through a single nodal Kirchhoff system; and (iv) a quantification of the pressure/flow sensitivity to the three-element Windkessel outlet, together with a transparent treatment of the whole-body outflow as an inflow-calibrated target (not a prediction) and a comparison of the untuned steady pressures against independent physiological data. 

\section{Methods}

\subsection{An overview of network anatomy and discretization}
\label{sec:anatomy}

The network is built on an anatomically derived graph of the human lymphatic system \cite{Savinkov2020,tretyakova2017developing}, comprising 1010 vertices and 1126 edges. We adapted this published graph under clinical-anatomist review to better represent regional lymph formation. For example, the hepatic territory, which carried no afferent inflow terminals in the original graph, was augmented with input leaves. The graph spans the body-wide afferent collectors (head and neck, upper and lower limbs, trunk wall, and the visceral beds that include hepatic, gastrointestinal, renal, splenic, and pulmonary), which converge through the lumbar and intestinal trunks into the cisterna chyli and then ascend the thoracic duct to the left lymphovenous junction (LVJ) \cite{Pan2017Atlas,Hematti2011}. Vertices carry one of three roles: 373 degree-one terminals (372 afferent inflow leaves plus one efferent outlet at the LVJ), 272 lymph-node vertices, and the remainder interior branch points. Each edge carries a length and an anatomical region label. The radius is not supplied by the anatomical data. Non-fixed edges enter the pipeline at a fallback $0.12$\,mm and the 11 thoracic-duct and cisterna-chyli seed edges are held fixed at $1.5$\,mm. The hydraulic radius that enters the solve is the output of the Murray closure of Sec.~\ref{sec:murray}, not this initialization. The region labels, which select the wall and valve parameters for each edge by our algorithms. %

\begin{figure}[t]
  \centering
  \includegraphics[width=0.9\linewidth]{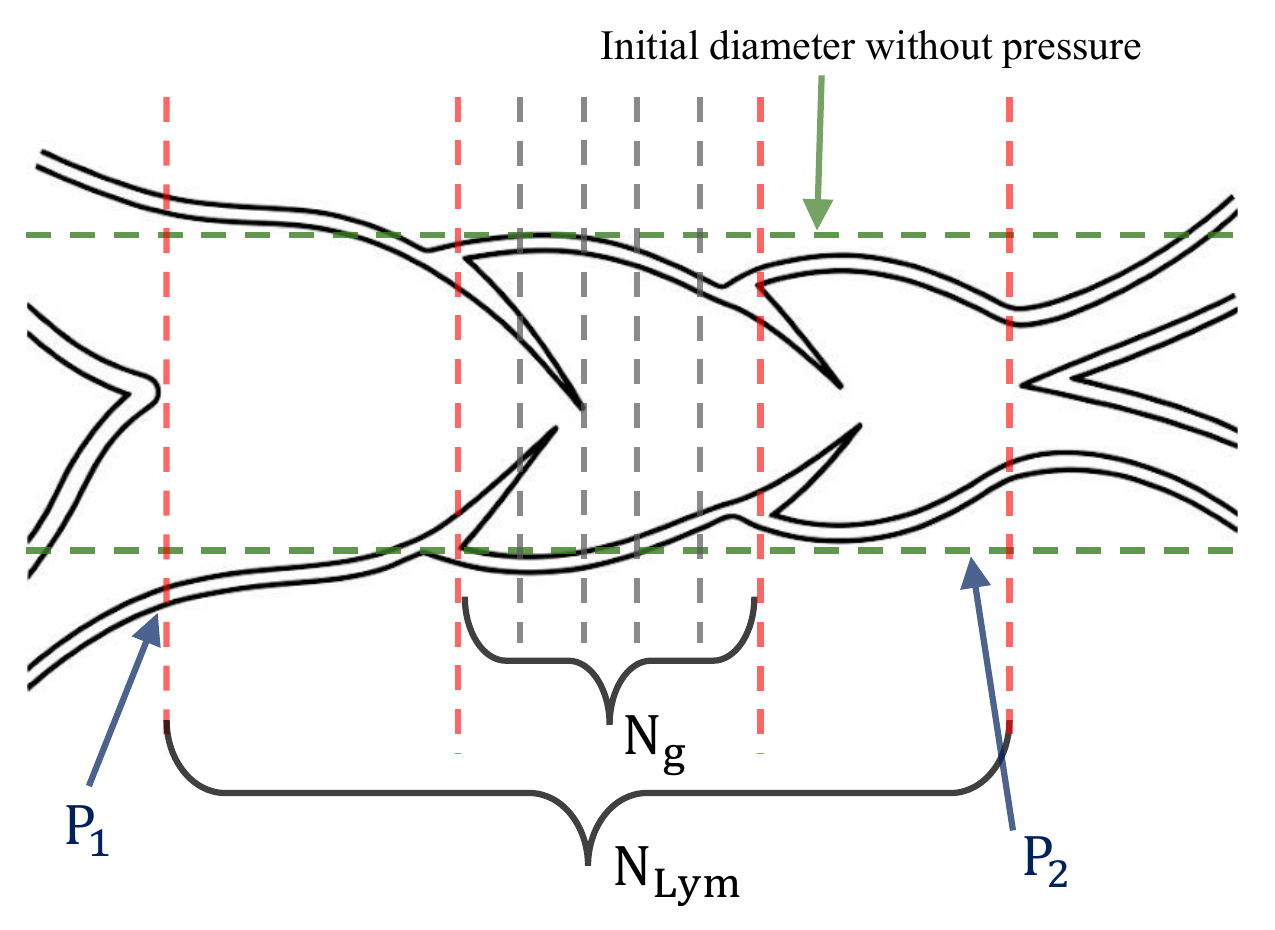}
  \caption{The model utilizes a three-level discretization: graph edges are discretized into $N_{\mathrm{lym}}$ sequential lymphangions (with interior vertices representing valve sites) as vertical red dashline, which are further discretized into $N_g$ wall sub-segments as vertical gray dashline. The horizontal green dashlines indicate vessel radius without pressure. Driven by the left-to-right flow, the pressure gradient ($P_1 > P_2$) naturally results in a slightly larger vessel radius at $P_1$ compared to $P_2$ within our simulation.}
  \label{fig:topology}
\end{figure}

To represent the discrete pumping units and their valves, each graph edge is expanded in two levels (Fig.~\ref{fig:topology}). First, an edge of length $L_{\mathrm{orig}}$ is divided into $N_{\mathrm{lym}}=\max\!\left(1,\operatorname{round}(L_{\mathrm{orig}}/s_{\mathrm{valve}})\right)$ series lymphangions, where the inter-valve spacing is $s_{\mathrm{valve}}=3$\,mm globally, at the upper end of the millimetre-scale lymphangion spacing reported for collecting lymphatics ($\sim$2-3\,mm in \cite{MargarisBlack2012}). The $N_{\mathrm{lym}}-1$ interior vertices created this way are the valve sites. Second, each lymphangion is divided into $N_g=5$ sub-segments, each carrying its own wall state but sharing the lymphangion phase. This allows the pressure distribution within a vessel to be non-linear, as opposed to the linear profile assumed in Hagen-Poiseuille flow, and permits the radius to vary along a single vessel. The production mesh contains 89{,}060 sub-edges.

The single global $s_{\mathrm{valve}}$ is a modelling choice, and the real spacing grows with vessel calibre. An ex vivo porcine study isolated thoracic-duct segments containing exactly one valve and found the valves separated by about $31$\,mm on average \cite{PatelLuKassab2025}, an order of magnitude above the peripheral collector value. Organ-resolved human spacing data are not available, so we keep $s_{\mathrm{valve}} = 3$ mm and uniform here.

Edge-level source terms, such as transmural leak in Eq.~(\ref{eq:leak}), are distributed equally to the two endpoint vertices. This symmetric split is mass conservative: after edge subdivision, adjacent half-contributions combine so that the total leak over each original edge is preserved exactly, independent of $N_{\mathrm{lym}}$ and $N_g$. Thus the discrete nodal mass balance in Eq.~(\ref{eq:massbalance}) is exactly mesh-independent, although this does not by itself imply pressure-field convergence. These source and sink terms are detailed in Secs.~\ref{sec:leak}--\ref{sec:bc}.

\subsection{Steady hydraulic network model}
\label{sec:hydraulic}

Lymph flow is treated as quasi-steady, laminar, and incompressible. For each edge $e=(i,j)$ the volumetric flow obeys Hagen--Poiseuille,
\begin{equation}
  Q_e = g_e\,(p_i-p_j), \qquad
  g_e = \frac{\pi r_e^{4}}{8\,\eta\,L_e\,(1+\phi_e)},
  \label{eq:poiseuille}
\end{equation}
where $p$ is the intraluminal nodal pressure (equal to the transmural pressure here, since we assume the external tissue pressure $P_{\mathrm{ext}}=0$ at this stage, Sec.~\ref{sec:leak}), $\eta = 1.8\times 10^{-3}$ Pa$\cdot$s is the lymph dynamic viscosity, $r_e$ and $L_e$ are the edge radius and length, the latter is edge-wisely fixed by original data from \cite{Savinkov2020}, and $\phi_e$ is the effective dimensionless valve resistance factor (Sec.~\ref{sec:valve}).

Eq.(\ref{eq:poiseuille}) is applied per sub-edge, not per anatomical vessel. Across the lumen we do not discretize, since the $r_e^{4}$ conductance is the closed-form solution of the axisymmetric cross-sectional flow problem, so the transverse direction enters analytically i.e.~keep $r_e$ as time-depended. Along the vessel axis we do discretize, into the sub-segment mesh of Sec.~\ref{sec:anatomy}, each sub-edge carrying its own radius and wall state. Each sub-edge is therefore approximately Poiseuille by construction (in a short time-step $dt$ in our simulation), while the vessel it belongs to is not, because its radius and pressure gradient change from one sub-edge to the next.

Mass conservation at every interior vertex $i$ yields the linear system as Kirchhoff equation
\begin{align}
  K\,\mathbf{p} & = \mathbf{b},\\
  K_{ii} & =\sum_{e\ni i} g_e + g^{\mathrm{leak}}_{i}
        + [G_{\mathrm{eff}}]_{i=996},\\
  K_{ij} & =-g_{e=(i,j)},
  \label{eq:kirchhoff}
\end{align}
with the right-hand side $\mathbf{b}$ accumulating the prescribed leaf inflows, the Starling-leak source, the lymph-node retention sink, and an ostial valve with the outlet Windkessel RCR source \cite{Westerhof1971,VignonClementel2006} (Secs.~\ref{sec:leak}--\ref{sec:bc}) as our boundary condition. $G_{\mathrm{eff}}$ is the contribution of that outlet to the diagonal, and it acts at the junction vertex alone (Sec.~\ref{sec:bc}). The lymph-node hydraulic resistance $R_{\mathrm{node}}$ enters through the conductances $g_e$ of node-incident edges (a series resistance),

\subsection{Murray radius closure: two independent derivations}
\label{sec:murray}

Pure fluid mechanics is insufficient to fully explain the $Q \propto r^3$ relationship for vessel calibers. Since standard Poiseuille flow yields $Q \propto r^4$ for a fixed pressure gradient, deriving the cubic scaling requires introducing an external physical mechanism. In this section, we reviewed two possible principles.

\textit{(a) Minimum-work principle }\cite{Murray1926}.
The total biological cost of a tube of radius $r$ carrying flow $Q$ is the sum of the viscous pumping power and a metabolic maintenance cost proportional to lumen volume,
\begin{equation}
  W_{\mathrm{total}}(r) = \frac{8\eta L Q^{2}}{\pi r^{4}} + b\,\pi r^{2} L.
  \label{eq:Wtot}
\end{equation}
Here $b$ is the metabolic cost per unit lumen volume and $L$ the tube length. Setting $\partial W_{\mathrm{total}}/\partial r=0$ gives $r^{6}=16\eta Q^{2}/(b\pi^{2})$, i.e. $Q\propto r^{3}$. Worth to note that if $b$ is radius-dependence, then the cube-law breaks. 

\textit{(b) Wall-shear-stress homeostasis.} 
For fully developed Poiseuille flow the wall shear stress is the kinematic identity
\begin{equation}
  \tau_w = \frac{4\eta Q}{\pi r^{3}}.
  \label{eq:tau}
\end{equation}
Endothelial cells regulate calibre to hold $\tau_w$ within a narrow homeostatic band ($\tau_w\!\equiv\!\tau_w^\star$) via mechanotransduction \cite{KamiyaTogawa1980,TarbellShi2010,Dixon2006}. Imposing $\tau_w=\tau_w^\star$ in Eq.~(\ref{eq:tau}) again gives $Q=(\pi\tau_w^\star/4\eta)\,r^{3}\propto r^{3}$ \cite{Sherman1981}. Both routes converge on the same cube exponent, which we adopt as $x_{\mathrm{Murray}}=3$.

\textit{Branching closure.}
At a bifurcation, mass conservation $Q_p=\sum_i Q_{c,i}$ combined with the per-branch form $Q_{c,i}=k\,r_{c,i}^{3}$, where the subscripts $p$ and $c$ denote the parent and the child branches and $k$ is a constant shared by the branches at that junction, yields both the aggregate cube law and
the explicit per-branch radius assignment
\begin{equation}
  r_p^{3} = \sum_{i} r_{c,i}^{3}, \qquad
  r_{c,i} = r_p\left(\frac{Q_{c,i}}{Q_p}\right)^{1/3}.
  \label{eq:branch}
\end{equation}
The flow-weighted form on the right resolves the under-determination of the aggregate rule (one equation, $n$ unknowns at an $n$-way branch). Radii are propagated from the thoracic-duct outlet (anatomically anchored at $r=1.5$\,mm) toward the leaves%
, and region-specific anatomical floors (i.e.~$0.5$\,mm for the abdominal/mesenteric and Hepatic trunks, $1.5$\,mm along the central thoracic duct). We adopt the classical $x_{\mathrm{Murray}}=3$ as the \emph{baseline} closure, applied along the central collectors and trunks where the cascade is anchored: these segments lie in the large-vessel regime in which the Newtonian--Poiseuille assumptions of derivations (a) and (b) are most accurate. We stress that $x_{\mathrm{Murray}}=3$ is a modelling closure, not a definitive biological law. For distal lymphatic collectors, the appropriate Murray-law exponent is not well constrained by direct experimental measurements. Existing lymphatic studies provide useful physiological context \cite{MargarisBlack2012,KornutaDixon2014}, but they do not establish a precise branching exponent for human lymphatic vessels. Thus, the single exponent used here should be read as a transparent baseline closure. The sensitivity of the network to $x_{\mathrm{Murray}}$ is examined in Sec.~\ref{sec:results}.

\subsection{Passive wall constitutive law}
\label{sec:wall}

Athanasiou et al.~\cite{Athanasiou2017} investigated, the passive mechanical properties of human collecting lymphatic vessels of pelvic origin. In that study, isolated\footnote{Explicitly eliminating both extrinsic tissue compression and intrinsic myogenic contraction.} vessels were cannulated and pressurized at varying levels of applied axial stretch $\ell/L$ in a calcium-free Krebs buffer, and the resulting pressure-diameter and force-elongation responses were found to be highly nonlinear. The vessels were subjected to transmural pressures from 0 to 25cm $\hbox{H}_2\hbox{O}$ at different axial stretch ratios, and the pressure-diameter behavior of the human vessels was reported to be very similar to data from rat mesenteric vessels \cite{Caulk2015, Razavi2017, Razavi2020}, despite the substantially larger vessel size. In the present work, we directly used the published pressure-diameter data \cite{Athanasiou2017} with axial stretch $\ell/L\in [1.00, 1.30]$ as the dataset for subsequent constitutive fitting and comparison.

To describe these measurements, we fitted the pressure-diameter data with a reparameterized passive relation written explicitly as pressure as a function of radius, 
\begin{equation}
  P_{\mathrm{passive}}(r) = \frac{T_{\mathrm{passive}}}{r} =
     k_1 \frac{\exp\!\big[k_2(\frac{r}{r_0}-1)\big]-1}{r},
  \label{eq:wall}
\end{equation}
where $T_{\mathrm{passive}}$ is the passive wall tension, $r_0 = 0.25\, r_{\max}$ is minimal vessel radius that provides transmural pressure, and the $r_{\max}$ is the maximum radius observed for each edge \cite{Athanasiou2017}. %
The material constants from the implicit three-parameter fit at an axial stretch of $\lambda_{\mathrm{axial}}=1.17$ are $k_1 = (1.6 \pm 1.5)\times 10^{-4}$ cmH$_2$O$\cdot$mm and $k_2 = 4.31 \pm 0.32$, with a joint $R^2 = 0.813$. It is worth noting that the data implies an edge dependence for $k_1$ and $k_2$. However, confirming this requires substantially more data. Therefore, we treat $k_1$ and $k_2$ as universal constants in this study.

In the network model, we calculate a distinct $r_0$ for each vessel. By inverting Eq.~(\ref{eq:wall}) using the baseline pressure and the Murray closure radius (Sec.~\ref{sec:murray}), we successfully anchor the exponential in the physiological mid-range. %

\subsection{Lymphatic valve mechanics}
\label{sec:valve}

\begin{figure}[t]
  \centering
  \includegraphics[width=1.03\linewidth]{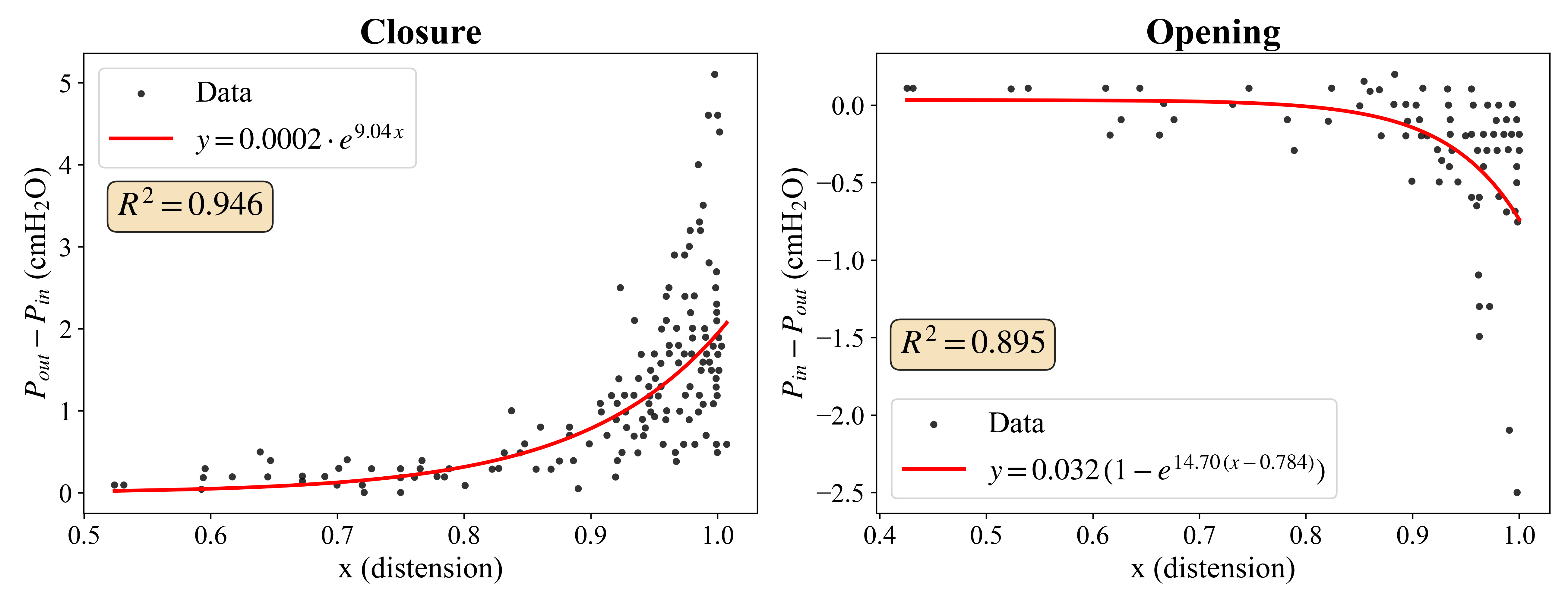}
  \caption{Pressure difference across the vessel versus distension $\texttt{x} \equiv r/r_{\max}$, with fitted curves (Eq.\ref{eq:vclose} and \ref{eq:vopen}). The left and right panels show the closure and opening processes, respectively, under Ca$^{2+}$-free conditions. Data are from Davis et al. (2011) \cite{Davis2011}. Due to raw data noise, the curves were fitted after median-binning the data into 15 bins. }
  \label{fig: PvsX}
\end{figure}

Recent studies have established a consistent physical picture for secondary lymphatic valves. Davis et al.~\cite{Davis2011} showed experimentally that rat mesenteric valves are biased toward the open state, exhibit opening--closure hysteresis, and require gating pressure differences that depend strongly on fractional distension $\texttt{x}\equiv r/r_{\max}$. Wilson et al.~\cite{Wilson2015} then constructed an idealized three-dimensional valve geometry from measured rat mesenteric valves. Building on these results, the Bertram et al, series \cite{Bertram2011,Bertram2014a,Bertram2014,Bertram2020,Bertram2023} (and related extension work \cite{Bertram2016,MooreBertram2018,ikhimwin2020computational}) introduced a sigmoid valve law, extended it to include transmural-pressure-dependent bias and hysteresis, and later refined the mechanics using three-dimensional fluid models. %

In this study, bicuspid secondary valves are modeled in a simplified manner using distension-dependent hysteretic thresholds, following \cite{Davis2011,Wilson2015,Bertram2014}.
Including $\Delta p=p_i-p_j$ as the signed pressure difference across the valve, the closing and opening thresholds are
\begin{align}
  \Delta p_{\mathrm{close}} &= a_p\exp(b_p \texttt{x}),
    \label{eq:vclose}\\
  \Delta p_{\mathrm{open}}  &= c_p\!\left[1-\exp\!\big(d_p(\texttt{x}-\texttt{x}_{\mathrm{valve},0})\big)\right],
    \label{eq:vopen}
\end{align}
with fitted variables as in Fig.~\ref{fig: PvsX},
\begin{align*}
    a_p^{(\mathrm{rat})} &= (2.30^{+3.49}_{-2.30}) \times 10^{-4} \;\; (\hbox{cmH}_2\hbox{O}),\\  b_p &= 9.04 \pm 1.12, \\
    c_p^{(\mathrm{rat})} &= (3.24^{+8.45}_{-3.24}) \times 10^{-2} \;\; (\hbox{cmH}_2\hbox{O}),\\  d_p &= 14.70 \pm 7.12, \\ \texttt{x}_{\mathrm{valve},0} &= 0.78 \pm 0.78,
\end{align*}
those the rat parameters $a_p^{(\mathrm{rat})}$ and $c_p^{(\mathrm{rat})}$ will be linearly rescaled to human pressure units in Sec.~\ref{sec:species}. The valve resistance enters Eq.~(\ref{eq:poiseuille}) through a dimensionless factor $\phi\in[\phi_{\mathrm{open}}=0,\ \phi_{\mathrm{closed}}=370]$, where the closed/open ratio $\approx 370$ is derived from the open- and closed-valve resistance ranges ($R_{Vx}/R_{Vn}$) measured in rat mesenteric lymphatics \cite{Bertram2014,Bertram2014a}. The discontinuous hysteretic switch is described by a state-dependent sigmoid,
\begin{equation}
  \phi(\Delta p,s) = \phi_{\mathrm{open}}
    + \frac{\phi_{\mathrm{closed}}-\phi_{\mathrm{open}}}
           {1+\exp\!\big[\sigma_{\mathrm{inv}}(\Delta p-\Delta p_{s})\big]},
  \label{eq:sigmoid}
\end{equation}
with the branch for valve state $s\in\{\mathrm{open},\mathrm{close}\}$ latched by a hysteresis dead-band. Because the thresholds describe the resistance of a whole lymphangion while our mesh subdivides each lymphangion into $N_g$ sub-segments, the per-valve resistance increment is distributed across the two adjacent gate sub-segments with a factor $N_g/2$ each (i.e. $\phi \to \phi N_g/2$), so that the series sum reproduces the intended lymphangion-level ratio as measured in \cite{Bertram2014,Bertram2014a}.

 The numerical parameter behavior as the following. The factor $\phi$ enters the edge conductance as $g=g_{\mathrm{tube}}/(1+\phi)$, which separates the conduit hydraulics $g_{\mathrm{tube}}\propto r^4/L$ from the dimensionless gating: $\phi_{\mathrm{open}}=0$ takes the fully-open, minimum-resistance state as reference, and the rat sigmoid steepness $\sigma_{\mathrm{inv}}^{(\mathrm{rat})}=0.1$\,Pa$^{-1}$ \cite{Bertram2014,Bertram2014a}. The threshold functions Eq.(\ref{eq:vclose}) and Eq.(\ref{eq:vopen}) are the only quantities fitted to the distension data \cite{Davis2011} as shown in Fig.~\ref{fig: PvsX}, with $R^2=0.95$ (closure) and $0.90$ (opening), respectively. In addition, the inter-valve spacing of human collecting lymphatics is on the order of $1$--$3$\,mm in the head and neck \cite{Pan2017Atlas,Pan2010}. We adopt 3 mm globally in this study.

\subsection{Transmural leak}
\label{sec:leak}

Transmural exchange is modelled as a linear hydraulic leak through the vessel wall. The volumetric leak from edge $e$ to a virtual interstitial ground at pressure $P_{\mathrm{ext}}$ is
\begin{equation}
  Q^{\mathrm{leak}}_e = g^{\mathrm{leak}}_e\,(p_e-P_{\mathrm{ext}}),
  \qquad
  g^{\mathrm{leak}}_e = L_{\mathrm{leak}}\,2\pi r_e L_e,
  \label{eq:leak}
\end{equation}
with $p_e=\tfrac12(p_i+p_j)$ the mean luminal pressure and the human filtration coefficient $L_{\mathrm{leak}}=1.25\times10^{-7}$\,cm\,s$^{-1}$\,cmH$_2$O$^{-1}$, obtained by allometric rescaling (Sec.~\ref{sec:species}) of the mouse baseline $17.3\times10^{-7}$ \cite{jannaway2021ve}. The value we use lies within one sigma of that rescaling, at the low end of the band, because a higher $L_{\mathrm{leak}}$ leaks away so much along the way that distal-limb lymph does not reach the thoracic duct. This is the hydraulic ($\Delta p$) term of the Starling relation, and the oncotic reflection term is not modelled separately but folded into the reference $P_{\mathrm{ext}}$, which is set to zero in this baseline so that intraluminal and transmural pressure coincide. %

\subsection{Lymph-node retention}
\label{sec:node}

Each lymph-node vertex $n$ is modeled as an effective sink that retains a fixed fraction $f_n = 10\%$~\cite{Adair1982_LymphProteinNodeTransit,Adair1983_LymphNodeVenousPressure} of the lymph flux entering the node. Since the retained flux depends on the instantaneous directed inflow, this sink is evaluated by a Picard fixed-point iteration. At each iteration, the directed inflow and retained flux are computed as
\begin{equation}
  Q^{\mathrm{in}}_n
  =
  \sum_{e\ni n}
  \frac{\left(p_{u(e)} - p_n\right)^{+}}{R_{\rm n}},
  \qquad
  Q^{\mathrm{ret}}_n
  =
  f_n Q^{\mathrm{in}}_n,
  \label{eq:ln_retention}
\end{equation}
where $u(e)$ denotes the endpoint of edge $e$ opposite node $n$, and $(x)^{+}=\max(x,0)$ ensures that only incident edges with flow entering the node contribute to $Q^{\mathrm{in}}_n$. The retained flux is then added to Kirchhoff Eq.(\ref{eq:kirchhoff}) as $b_n \mathrel{-}= Q^{\mathrm{ret}}_n$.

We do not assume a universal relationship between lymph-node volume and hydraulic resistance. Instead, the effective nodal resistance entering $g_e$ is based on the apparent mouse popliteal-node resistance reported by \cite{Jafarnejad2015},
\begin{equation}
  R_{\mathrm{n}}^{(\mathrm{mouse})}
\simeq 6.8~\mathrm{mmHg}/(\mu\mathrm{L}/\mathrm{min}).
  \label{eq:mouse_node_resistance}
\end{equation}
In the absence of region-specific human hydraulic data, we assign a common baseline effective resistance to all lymph-nodes. But a more accurate human size-dependent value should be updated in future experiments, see also \cite{Jafarnejad2015,Nagai2008_LymphNodeInflammation}.

\subsection{Boundary conditions}
\label{sec:bc}

Inflow enters as a Neumann (prescribed-flux) condition. Each anatomical region $\mathcal{R}$ is assigned a literature-based fractional contribution $f_{\mathcal{R}}=Q_{\mathcal{R}}/Q_{\mathrm{TD}}$ to the total thoracic-duct lymph flow. These regional priors (Table~\ref{tab:regional_td_lymph_fraction}) are compiled from the classical human postnodal distribution \cite{YoffeyCourtice1970,Levick1991,Morris1956} together with quantitative animal tracer studies \cite{Mobley1989,Steil1993,Li2011ThoracicDuctLymphRat,Semaeva2010SpleenLymph}, with the hepatic, gastrointestinal, and renal beds dominating. The regional load $Q_{\mathcal{R}}$ is then distributed over that region's $n_{\mathcal{R}}$ degree-one afferent leaves,
\begin{equation}
  Q_{i} = \frac{Q_{\mathcal{R}}}{n_{\mathcal{R}}}\,w_i, \qquad w_i \ge 0,
  \label{eq:qin}
\end{equation}
where the within-region weights $w_i$ are not held uniform: a subset of high-capacity collecting leaves carry weights $>1$ (and some peripheral leaves $<1$), reflecting heterogeneous per-vessel drainage and chosen to suppress nonphysical counterflow (lymph entering a terminal afferent leaf). %

Table~\ref{tab:inflow} reports the realized, as-simulated inflows that the network actually receives. The \emph{relative} regional rates $Q_{\mathcal{R}}$ (Table~\ref{tab:inflow}) follow the classical regional proportions of lymph production e.g.~hepatic and gastrointestinal beds dominating \cite{YoffeyCourtice1970,Levick1991}, and we calibrate it so that the resulting net thoracic-duct outflow falls in the physiological range. With this calibration the gross load is $\sum_{\mathcal{R}} Q_{\mathcal{R}} = 65.75$\,mm$^3$/s ($=5.68$\,L/day), which is consistent with classical estimates of net interstitial filtration entering the lymphatics ($\sim$8\,L/day) \cite{LevickMichel2010}.

Mass conservation at steady state relates the (calibrated) inflow to the output via transmural reabsorption and nodal retention,
\begin{equation}
  \underbrace{\sum_{i}Q_i}_{\text{gross inflow}}
  = \Qout
  + \underbrace{\sum_{e} Q^{\mathrm{leak}}_{e}}_{\text{Starling reabsorption}}
  + \underbrace{\sum_{n\in\mathrm{LN}} Q^{\mathrm{ret}}_{n}}_{\text{nodal retention}},
  \label{eq:massbalance}
\end{equation}
so $\Qout$ is computed from the network solve (not pinned at the outlet), but its \emph{scale} is fixed by the inflow calibration above (Sec.~\ref{sec:results}).

In addition, a respiration-driven pressure difference of $\Delta P\approx 15$\,cmH$_2$O ($\approx 1.5$\,kPa, split as $\sim$7.5\,cmH$_2$O across each of its two trunk segments) is imposed along the cisterna-chyli trunk, representing the extrinsic respiratory pump that drives central lymph return as the descending diaphragm modulates intra-abdominal pressure over the sub-diaphragmatic cistern \cite{Moazzam2026,kirkpatrick2013intra}. The magnitude sits inside the respiratory swing measured in diaphragmatic collecting lymphatics, where the intraluminal pressure moves between about $-20$ and $+10$\,cmH$_2$O around a $4$\,cmH$_2$O end-expiratory value \cite{Solari2020,Moriondo2016}.

\begin{table}[t]
  \centering
    \caption{
    Regional inflow boundary conditions. \emph{Delivered} is the path-specific
    fraction of each regional inflow reaching the outlet, from a linear
    flux-transport solve on the converged directed network.
    }
  \label{tab:inflow}
  \small
  \begin{tabular}{lrrr}
    \toprule
    Inlet region & $Q_{\mathcal{R}}$ & Delivered & Outlet \\
                 & (mm$^3$/s)        & (\%)                  & (\%)                      \\
    \midrule
    GI/mesenteric collector              & 19.553 & 53.2   & 31.2   \\
    Hepatic interstitium                 & 16.413 & 72.4   & 35.6   \\
    Lower limb, foot dorsum              & 11.067 & 5.0    & 1.7    \\
    Renal interstitium                   &  4.220 & 64.5   & 8.2    \\
    Upper limb, superficial collector    &  2.315 & 53.5   & 3.7    \\
    Pulmonary/bronchomediastinal         &  2.000 & 75.9   & 4.5    \\
    Upper limb, fingertip                &  1.853 & 69.7   & 3.9    \\
    Spleen                               &  1.694 & 73.2   & 3.7    \\
    Upper limb, axillary                 &  1.544 & 63.4   & 2.9    \\
    Head/neck, superficial cervical      &  1.394 & 30.0   & 1.3    \\
    Lower limb, pelvic region            &  0.927 & 17.5   & 0.5    \\
    Lower limb, deep collector           &  0.617 & 8.3    & 0.2    \\
    Pulmonary lung parenchyma            &  0.591 & 60.4   & 1.1    \\
    Upper limb, deep collector           &  0.539 & 43.2   & 0.7    \\
    Lower limb, superficial collector    &  0.420 & 8.0    & 0.1 \\
    Thoracic trunk, subclavian           &  0.302 & 73.9   & 0.7    \\
    Lower limb, ankle superficial        &  0.184 & 3.6    & $<0.1$ \\
    Thoracic wall, parasternal           &  0.093 & 33.1   & $<0.1$ \\
    Head/neck, deep cervical             &  0.023 & 82.1   & $<0.1$ \\
    \midrule
    Total                                & 65.750 & 50.7   & 100.0  \\
    \bottomrule
  \end{tabular}
\end{table}

\begin{table*}[t]
\centering
\small
\caption{
Working priors for regional attribution of thoracic-duct lymph outflow. Entries
are approximate attribution priors (old human postnodal estimates plus animal
consistency checks), not measured organ-level production fractions; unresolved
abdominal and pelvic sources are treated as residual.
}
\label{tab:regional_td_lymph_fraction}
\resizebox{\textwidth}{!}{%
\begin{tabular}{cccc}
\hline
Region/source & Reported contribution to $Q_{\rm TD,out}$ & Adopted prior for $f_{\mathcal{R}}$ & Evidence / note \\
\hline
\makecell[c]{Liver /\\ hepatic lymph}
&
\makecell[c]{$30$--$49\%$ in human\\ postnodal TD estimates;\\ $\sim 1/3$ of resting TD flow\\ in dogs}
&
\makecell[c]{$0.35$--$0.45$;\\ nominal $0.40$}
&
\makecell[c]{Directly supported by the human\\ postnodal TD table\\ \cite{YoffeyCourtice1970,Levick1991};\\ dog hepatic-occlusion consistency\\ check \cite{Mobley1989}.}
\\
\hline
\makecell[c]{Gastrointestinal / mesenteric /\\ extrahepatic portal bed}
&
\makecell[c]{$\sim 37\%$ in human\\ postnodal TD estimates;\\ total splanchnic contribution\\ $54\pm5$--$78\pm5\%$ in dogs}
&
\makecell[c]{$0.25$--$0.38$;\\ nominal $0.32$}
&
\makecell[c]{Human table supports the GI entry\\ \cite{YoffeyCourtice1970,Levick1991};\\ dog tracer data constrain total\\ splanchnic, not isolated GI, flow\\ \cite{Steil1993}.}
\\
\hline
\makecell[c]{Kidney /\\ renal lymph}
&
\makecell[c]{$6$--$11\%$ in human\\ postnodal TD estimates}
&
\makecell[c]{$0.06$--$0.10$;\\ nominal $0.08$}
&
\makecell[c]{Directly supported by the human\\ postnodal TD table, but still\\ an approximate old estimate\\ \cite{YoffeyCourtice1970,Levick1991}.}
\\
\hline
\makecell[c]{Lung /\\ pulmonary lymph}
&
\makecell[c]{$3$--$15\%$ in human\\ postnodal TD estimates}
&
\makecell[c]{$0.03$--$0.10$;\\ nominal $0.06$}
&
\makecell[c]{Directly supported by the human\\ postnodal TD table; large\\ physiologic variability is expected\\ \cite{YoffeyCourtice1970,Levick1991}.}
\\
\hline
\makecell[c]{Spleen}
&
\makecell[c]{Not directly reported\\ as a fraction of human\\ TD outflow}
&
\makecell[c]{$0.00$--$0.08$;\\ nominal $0.04$}
&
\makecell[c]{Rat spleen efferent lymph can be\\ cannulated and measured, but this\\ does not directly give a spleen/TD\\ outflow fraction \cite{Semaeva2010SpleenLymph}.\\ Treated as a weak, indirect prior.}
\\
\hline
\makecell[c]{Limbs, cervical trunks,\\ and body-wall residual}
&
\makecell[c]{$<10\%$ for limbs and\\ cervical trunks in the human table;\\ non-splanchnic contribution can be\\ larger in dog tracer studies}
&
\makecell[c]{residual, $0.00$--$0.31$;\\ nominal $0.09$}
&
\makecell[c]{Human table supports a small\\ limb/cervical contribution\\ \cite{YoffeyCourtice1970,Levick1991};\\ dog tracer data are used only as\\ non-human consistency constraints\\ \cite{Steil1993}.}
\\
\hline
\makecell[c]{Other abdominal /\\ pelvic organs}
&
\makecell[c]{Not cleanly\\ separated}
&
\makecell[c]{residual;\\ nominal $0.01$}
&
\makecell[c]{Not independently constrained.\\ Used only to close the attribution\\ budget after the better-supported\\ regional priors are assigned.}
\\
\hline
\end{tabular}%
}
\end{table*}

The thoracic-duct outlet is the LVJ, where the terminal duct drains into the left subclavian--internal-jugular venous angle \cite{Ratnayake2018,Petrenko2004}. We terminate it with a three-element (RCR) Windkessel \cite{Westerhof1971,VignonClementel2006}, of impedance $Z(\omega)=R_p+R_d/(1+\mathrm{i}\,\omega R_d C)$, where $\omega$ is the angular frequency. Because the present model is steady the compliance carries no current. We take the direct-current limit $C\to0$, in which the outlet reduces to a series resistance $R_{\mathrm{LVJ}}=R_p+R_d$ and avoids a hard Dirichlet pin,
\begin{align}
  \notag Q_{\mathrm{out}} &= g_{\mathrm{out}}\,(p_{\rm out}-P_{\mathrm{subclav}}),\\
  g_{\mathrm{out}} &=\frac{1}{R_p+R_d}=\frac{1}{R_{\mathrm{LVJ}}},
  \label{eq:rcr}
\end{align}
with $P_{\mathrm{subclav}}=3$\,mmHg ($400$\,Pa) the supine-resting central-venous far field \cite{lloyd2025measurement}. That review gives a mean central venous pressure of $2$--$3$\,mmHg in healthy awake patients, and notes that in the supine position there is little pressure difference between the intrathoracic central veins and the right atrium. The $8$--$12$\,mmHg often quoted as the normal range \cite{shah2018physiology} is a catheter-referenced value. It is read in ventilated and critically ill patients, where raised intrathoracic pressure and reduced right-ventricular compliance lift the reading \cite{lloyd2025measurement}.
We set $C=10^{-15}$\,m$^{3}$\,Pa$^{-1}$ to enforce this direct-current limit numerically.\footnote{The steady update relaxes the reservoir by implicit Euler: with $\beta=\Delta t/C$ and $D_{\mathrm{w}}=1+\beta/R_p+\beta/R_d$ it adds $G_{\mathrm{eff}}=(1+\beta/R_d)/(R_p D_{\mathrm{w}})$ to the outlet diagonal of the conductance matrix, and $G_{\mathrm{eff}}\to 1/(R_p+R_d)=g_{\mathrm{out}}$ as $\beta\to\infty$ ($C\to0$), recovering Eq.~(\ref{eq:rcr}).} A physiological compliance would instead drive $G_{\mathrm{eff}}\to1/R_p$, dropping $R_d$ and underestimating the junction pressure. The proximal/distal resistance split and the compliance become active only in the pulsatile, contraction-driven companion study.

We do not fit $R_{\mathrm{LVJ}}$. We anchor it to the measured junction pressure. Lu \emph{et al.} reported a thoracic-duct pressure of $8.2\pm3.2$\,mmHg at the LVJ \cite{Lu2022}, although this is a porcine measurement. A directly comparable human junction pressure is not firmly established: Kelly \emph{et al.} measured a higher mean intraluminal thoracic-duct pressure of $16\pm5$\,mmHg \cite{Kelly2022}, but in patients receiving positive-pressure mechanical ventilation, whereas anatomical-physiological studies of the terminal thoracic duct and its ostial valve adopt a comparable few-mmHg outlet pressure at the duct-to-vein junction \cite{Ratnayake2018}. With the calibrated net outflow, an overbar denoting the calibrated operating value, this gives
\begin{align}
  R_{\mathrm{LVJ}} & =\frac{\bar p_{\rm out}-P_{\mathrm{subclav}}}{\bar Q_{\mathrm{out}}}\approx1.96\times10^{10}\ \mathrm{Pa\,s\,m^{-3}}\label{eq:rlvj}\\
  \notag R_p & =4.50\times10^{9},\ R_d=1.51\times10^{10},
\end{align}
Because $g_{\mathrm{out}}$ is held fixed, the junction pressure is an emergent linear function of outflow (slope $R_{\mathrm{LVJ}}$, offset $P_{\mathrm{subclav}}$): across the physiological delivery range ($2.0$--$3.0$\,L/day) \cite{GuytonHall2015,Moazzam2026} it spans roughly $6$--$9.5$\,mmHg, increasing monotonically with flow and remaining within the clinical band of \cite{Lu2022}. We then calibrate $\bar p_{\rm out}$. The outlet sensitivity to its parameters is reported in Sec.~\ref{sec:results}.

\subsection{Allometric species rescaling}
\label{sec:species} 

To apply rodent-derived valve mechanics to the human thoracic duct, we use a
scale-invariant mapping. The assumption is limited but explicit: rodent and
human collecting lymphatics share the same qualitative valve behavior,
including opening bias, hysteresis, and distension-dependent opening and
closure thresholds. The absolute pressure, diameter, and length scales are then
replaced by human thoracic-duct scales. We therefore define%
\begin{align}
\label{Eq: rescale}
\notag \alpha_p &\equiv \frac{P_{\mathrm{human}}}{P_{\mathrm{rat}}}
= 9.10 \pm 2.90
\quad \text{(measurement-based)}, \\
\notag
\lambda_r &\equiv \frac{d_{\mathrm{human}}}{d_{\mathrm{rat}}}
= 4.34 \pm 1.73
\quad \text{(surrogate-based)}, \\
\notag
\lambda_L &\equiv \frac{L_{\mathrm{human}}}{L_{\mathrm{rat}}}
= 7.97 \pm 1.87
\quad \text{(range-derived)}, \\
\notag
\alpha_p^m &\equiv \frac{P_{\mathrm{human}}}{P_{\mathrm{mouse}}}
= 3.30 \pm 1.90
\quad \text{(outlet-pressure proxy)}, \\
\notag
\lambda_r^m &\equiv \frac{d_{\mathrm{human}}}{d_{\mathrm{mouse}}}
= 4.48 \pm 1.77
\quad \text{(surrogate-based)}, \\
\lambda_L^m &\equiv \frac{L_{\mathrm{human}}}{L_{\mathrm{mouse}}}
= 11.9 \pm 4.7
\quad \text{(range-derived)},
\end{align}
where $P_{\mathrm{human}}\simeq 8\pm4~\mathrm{mmHg}$ is taken from thoracic-duct pressure measurements at the lympho-venous junction. Its central value follows the $8.2\pm3.2$\,mmHg control group of Lu \emph{et al.}~\cite{Lu2022}, which was measured in domestic pigs, so the human side of $\alpha_p$ rests on the same porcine anchor as $R_{\mathrm{LVJ}}$ in Sec.~\ref{sec:bc}. The direct human recording, $16\pm5$\,mmHg from Kelly \emph{et al.}~\cite{Kelly2022}, is higher and was made under positive-pressure mechanical ventilation. Because these measurements were obtained under clinical conditions and may depend on ventilation and measurement setting, we use them only as a broad human pressure scale. The rat pressure scale $P_{\mathrm{rat}}$ is estimated from the operating-pressure range reported in rat thoracic-duct contractility studies~\cite{Gasheva2017}.

The diameter scale uses $d_{\mathrm{human}}=2.69\pm1.06~\mathrm{mm}$ from resting human thoracic-duct sonography~\cite{Shinaoka2025} and $d_{\mathrm{rat}}=0.620\pm0.035~\mathrm{mm}$ from isolated rat thoracic-duct measurements at controlled transmural pressure~\cite{Gasheva2017}. The length scale uses human thoracic-duct anatomical data from 30 males and 15 females ~\cite{HumanTDLength2013} and reported rat thoracic-duct anatomical ranges ~\cite{Li2011ThoracicDuctLymphRat}.

The mouse-to-human scaling factors in Eq.~\eqref{Eq: rescale} are estimated in the same way, mouse collecting-lymphatic radius and pressure priors from Davis et al.~\cite{Davis2026_MouseLymphaticPressure}. They did not directly measure the mouse thoracic duct. Their servo-null micropuncture data were obtained from other mouse lymphatic networks, including cervical, popliteal, and mesenteric lymphatics. %

These ratios propagate to each parameter by its physical dimension. A conduit conductance follows Poiseuille scaling, so a rodent segment maps to human for  the lymph-node resistance as 
\begin{align}
R_{\mathrm{node}}^{\mathrm{human}}=R_{\mathrm{node}}^{\mathrm{mouse}}\,\lambda_L^m/(\lambda_r^m)^{4}.   
\end{align}
The valve pressure thresholds Eq.(\ref{eq:vclose}) and Eq.(\ref{eq:vopen}) scale by $\alpha_p$ and the sigmoid steepness inversely, 
\begin{align}
\sigma_{\mathrm{inv}}^{\mathrm{human}}=\sigma_{\mathrm{inv}}^{\mathrm{rat}}/\alpha_p.
\end{align}
The dimensionless valve shape parameters ($b_p,d_p,\texttt{x}_{\mathrm{valve},0}$) are unchanged. The wall filtration coefficient scales as 
\begin{align}
    L_{\mathrm{leak}}^{\mathrm{human}}=L_{\mathrm{leak}}^{\mathrm{mouse}}(\lambda_r^m)^{3}/(\lambda_L^m)^{2}.
\end{align}

\subsection{Cycle-space radius perturbation for pressure relaxation}
\label{sec:perturbation}

At the calibrated delivery the raw passive solve is globally over-pressured
(Sec.~\ref{sec:results}). Because the Murray closure (Sec.~\ref{sec:murray})
fixes the radii from a first flow estimate, the over-pressured edges are simply
too narrow for the load they carry. We relieve the over-pressure by selectively
widening these edges while holding the prescribed boundary fluxes
$\mathbf{q}_{\mathrm{ext}}$ fixed, so that the net outflow and the regional
inflows are preserved.

Represent the directed network by a signed vertex--edge incidence matrix
$\mathbf{A}\in\mathbb{R}^{n\times m}$ over its $n$ vertices and $m$ edges, and let
$\mathbf{C}\in\mathbb{R}^{L\times m}$ be the fundamental cycle--edge matrix: each
of its $L=m-n+1$ rows is one independent loop of a fundamental cycle basis, with
$\mathbf{C}_{ij}=\pm1$ when edge $j$ lies in loop $i$ (the sign set by whether
the edge and loop orientations agree) and $0$ otherwise. Let
$\mathbf{Q}\in\mathbb{R}^m$ be the signed edge flows and
$\mathbf{R}=\operatorname{diag}(R_e)$ the diagonal matrix of Hagen--Poiseuille
edge resistances, $R_e=8\eta L_e/(\pi r_e^{4})=1/g_e$. At fixed boundary flux, a
particular flow $\mathbf{Q}_0$ obeying nodal mass conservation
$\mathbf{A}\mathbf{Q}_0=\mathbf{q}_{\mathrm{ext}}$, plus a cycle-space correction
$\mathbf{Q}=\mathbf{Q}_0+\mathbf{C}^\top\mathbf{z}$ in the loop currents
$\mathbf{z}\in\mathbb{R}^{L}$, solves the steady problem through Kirchhoff's loop
law $(\mathbf{C}\mathbf{R}\mathbf{C}^\top)\mathbf{z}=-\mathbf{C}\mathbf{R}\mathbf{Q}_0$.

Perturbing the radii, $\mathbf{R}\to\mathbf{R}+\Delta\mathbf{R}$ at fixed
$\mathbf{q}_{\mathrm{ext}}$, and linearizing in $\Delta\mathbf{R}$ gives the
cycle-projected flow response
\begin{equation}
  \Delta \mathbf{Q} = -\,\mathbf{C}^\top\,(\mathbf{C}\mathbf{R}\mathbf{C}^\top)^{-1}\,\mathbf{C}\,\Delta\mathbf{R}\,\mathbf{Q},
  \label{eq:dQ_perturb}
\end{equation}
where $\Delta\mathbf{R}=\operatorname{diag}(\Delta R_e)$ is the diagonal
resistance perturbation. The operator
$\mathbf{C}^\top(\mathbf{C}\mathbf{R}\mathbf{C}^\top)^{-1}\mathbf{C}$ projects a
per-edge perturbation back onto the cycle space, enforcing mass conservation and
loop balance. The nodal pressures follow from the updated flows and resistances. Equation~(\ref{eq:dQ_perturb}) is applied as a damped, widen-only relaxation that enlarges only the over-pressured edges in bounded substeps, holding the per-edge flow drift within tolerance, so the field relaxes toward the physiological range without altering the calibrated outflow $Q_{\mathrm{out}}$. Three conditions stop the relaxation. The
$95$th percentile of $|p|$ must fall to $30$\,mmHg, the maximum to
$50$\,mmHg, and the $99$th percentile of the forward-edge valve factor $\phi$
below $10$. We set the pressure levels below the clinical cuff-occlusion
ceilings of Sec.~\ref{sec:validation}, which reach $39\pm14$\,mmHg in the upper
limb \cite{Modi2007} and $25.2\pm16.7$\,mmHg in the lower leg \cite{Unno2010},
so that a field meeting them sits inside the measured range. %

\subsection{Solution procedure}
\label{sec:pipeline}

\begin{figure*}[t]
  \centering
  \begin{tikzpicture}[
    >={Latex[length=2.2mm,width=1.5mm]},
    node distance=7mm and 8mm,
    stage/.style={draw, rounded corners=2pt, align=center,
                  text width=38mm, minimum height=20mm, inner sep=3pt,
                  font=\footnotesize},
    input/.style={stage, fill=blue!5, draw=blue!50!black},
    solve/.style={stage, fill=green!6, draw=green!45!black},
    result/.style={stage, fill=orange!10, draw=orange!65!black},
    focus/.style={stage, fill=orange!18, draw=orange!80!black, very thick},
    valid/.style={stage, fill=gray!10, draw=black!45},
    flow/.style={->, line width=0.55pt, draw=black!70, rounded corners=4pt},
    loop/.style={->, line width=0.6pt, draw=orange!75!black, dashed, rounded corners=4pt}
  ]
    \node[input] (s1) {\textbf{(1) Graph geometry}\\[1pt]
      {\scriptsize clean graph; every organ\\carries afferent inputs;\\label organ regions}\\[-1pt]
      {\scriptsize length fixed; $r$ init $0.12$\,mm}\\[-1pt]
      {\scriptsize Sec.~\ref{sec:anatomy}}};

    \node[input, right=of s1] (s2) {\textbf{(2) Organ-wise}\\\textbf{inflow $Q_{\mathrm{in}}$}\\[1pt]
      {\scriptsize per-leaf $Q_{\mathrm{in}}$ from\\regional TD fractions;\\scale set by fixing $Q_{\mathrm{out}}$}\\[-1pt]
      {\scriptsize Table~\ref{tab:regional_td_lymph_fraction}, Sec.~\ref{sec:bc}}};

    \node[solve, right=of s2] (s3) {\textbf{(3) Flow solve +}\\\textbf{Murray radii}\\[1pt]
      {\scriptsize solve $Q,P$ (RCR outlet);\\TD radius anchored $1.5$\,mm;\\$r_p^3=\sum_i r_{c,i}^3$ (Eq.~\ref{eq:branch})}\\[-1pt]
      {\scriptsize Secs.~\ref{sec:murray},\ref{sec:bc}}};

    \node[solve, below=of s3] (s4) {\textbf{(4) Wall law +}\\\textbf{anchor homotopy}\\[1pt]
      {\scriptsize apply Eq.~(\ref{eq:wall});\\edge-wise $P_{\mathrm{anchor}}$ aids\\convergence, $P_{\mathrm{anchor}}\!\to\!0$}\\[-1pt]
      {\scriptsize Sec.~\ref{sec:wall}}};

    \node[solve, left=of s4] (s5) {\textbf{(5) Cycle-space}\\\textbf{radius widening}\\[1pt]
      {\scriptsize widen over-pressured edges\\under bounded per-edge\\flow drift (Eq.~\ref{eq:dQ_perturb})}\\[-1pt]
      {\scriptsize Sec.~\ref{sec:perturbation}}};

    \node[solve, left=of s5] (s6) {\textbf{(6) Expand +}\\\textbf{valves + leak/node}\\[1pt]
      {\scriptsize $N_{\mathrm{lym}}$ lymphangions, $N_g{=}5$;\\Davis/Bertram valves; $K\mathbf{p}{=}\mathbf{b}$\\with leak + node retention;\\$Q_{\mathrm{in}}$ re-calibrated}\\[-1pt]
      {\scriptsize Secs.~\ref{sec:valve},\ref{sec:leak},\ref{sec:node}}};

    \node[solve, below=of s6] (s7) {\textbf{(7) Counterflow}\\\textbf{suppression}\\[1pt]
      {\scriptsize cycle-space block analysis;\\re-update feeding-leaf $Q_{\mathrm{in}}$\\(and local radii)}\\[-1pt]
      {\scriptsize Sec.~\ref{sec:bc}}};

    \node[focus, right=of s7] (s8) {\textbf{(8) Self-consistent}\\\textbf{pressure relaxation}\\[1pt]
      {\scriptsize iterate (5)--(7): widen +\\re-anchor $P_{\mathrm{anchor}}\!\to\!0$,\\re-update $Q_{\mathrm{in}}$ until $p$ stalls\\($r\!\in\![0.05,1.5]$\,mm exc.\ CC/TD,\\$p_{\max}\!\le\!50$, $Q_{\mathrm{out}}$ fixed)}\\[-1pt]
      {\scriptsize median $\sim$32\,mmHg; Sec.~\ref{sec:res-relax}}};

    \node[valid, right=of s8] (s10) {\textbf{(9) Validation /}\\\textbf{sensitivity}\\[1pt]
      {\scriptsize pressures vs clinical data;\\sweeps over $r_{\rm floor}$,\\$x_{\mathrm{Murray}}$, outlet impedance}\\[-1pt]
      {\scriptsize Secs.~\ref{sec:validation},\ref{sec:results}}};

    \draw[flow] (s1) -- (s2);
    \draw[flow] (s2) -- (s3);
    \draw[flow] (s3.east) -- ++(4mm,0) |- (s4.east);
    \draw[flow] (s4) -- (s5);
    \draw[flow] (s5) -- (s6);
    \draw[flow] (s6.west) -- ++(-4mm,0) |- (s7.west);
    \draw[flow] (s7) -- (s8);
    \draw[loop] (s8.north) -- node[right=1pt, font=\scriptsize, text=orange!50!black]
      {iterate (5)--(7)} (s5.south);
    \draw[flow] (s8) -- (s10);
  \end{tikzpicture}
  \caption{Static solution procedure (Sec.~\ref{sec:pipeline}). Steps (1)--(4)
  build and close the network; steps (5)--(7) form the outer loop, iterated by the
  self-consistent pressure relaxation (8) (dashed return) until the pressure field
  stalls; step (9) is validation and sensitivity. Step (8) is the central result.}
  \label{fig:pipeline}
\end{figure*}
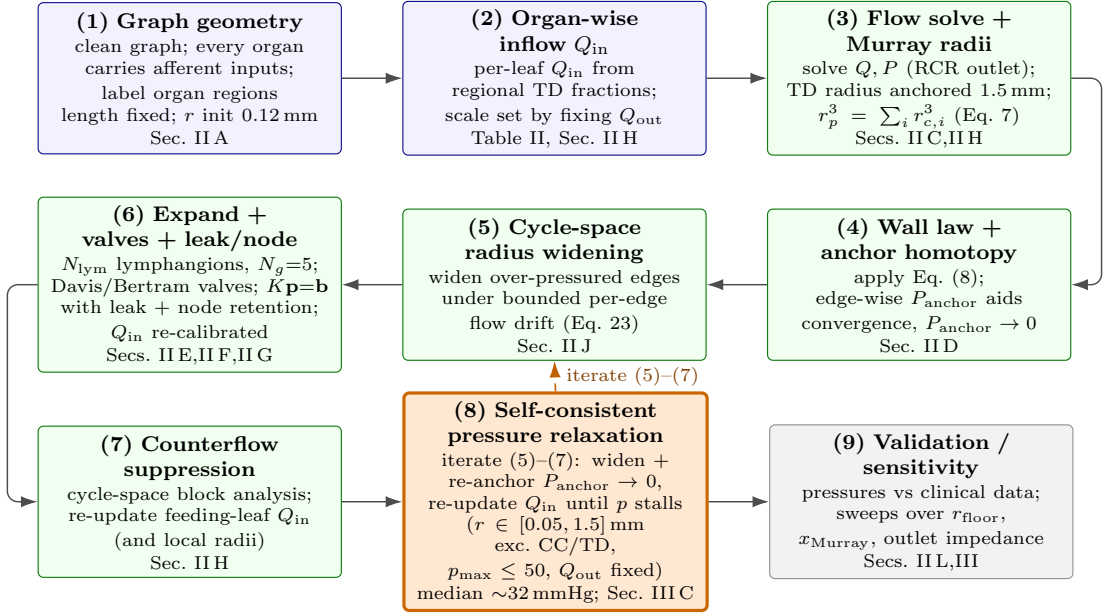

The static model is assembled in a fixed sequence (Fig.~\ref{fig:pipeline}): \\
\textbf{(1)} clean the
geometry of the lymphatic graph, ensure that every organ territory carries
afferent inputs, and label the organ regions. Each edge length is fixed by the
human anatomical data, while the edge radius is a priori unknown and is
initialized to a fallback $0.12$\,mm, except for the 11 fixed thoracic-duct and cisterna-chyli seed edges at $1.5$\,mm (Sec.~\ref{sec:anatomy});\\
\textbf{(2)} assign an organ-wise inflow $Q_{\mathrm{in}}$ to every afferent leaf
following the regional thoracic-duct fractions
(Table~\ref{tab:regional_td_lymph_fraction}), with the overall scale set by
fixing the target outflow $Q_{\mathrm{out}}$ so that each leaf initially carries
a relatively small inflow (Sec.~\ref{sec:bc});\\
\textbf{(3)} solve the network for the per-edge flow $Q$ and pressure $P$, closing the
system with the three-element (RCR) venous outlet (Sec.~\ref{sec:bc}). Then,
anchoring the thoracic-duct radius at $1.5$\,mm, reassign every edge radius
by Murray's law and Eq.~(\ref{eq:branch}) from the resulting flows
(Sec.~\ref{sec:murray}); \\
\textbf{(4)} apply the passive wall law, Eq.~(\ref{eq:wall}). Because its exponential
form makes the radius solve numerically stiff, we add an edge-wise constant
anchor pressure $P_{\mathrm{anchor}}$ to Eq.~(\ref{eq:wall}) to aid convergence
and finally relax $P_{\mathrm{anchor}}\to 0$ (Sec.~\ref{sec:wall}); \\
\textbf{(5)} jointly adjust the vessel radii and the pressure field with the steady
hydraulic perturbation analysis of Sec.~\ref{sec:perturbation}, a Kirchhoff
cycle-space linearization that widens the over-pressured edges under a bounded
per-edge flow-drift constraint;\\
\textbf{(6)} expand the graph into lymphangions and sub-segments, activate the valves,
and re-solve the Kirchhoff system with the wall law, now including the
transmural leak and lymph-node retention as source/sink terms in
$K\mathbf{p}=\mathbf{b}$. The added sinks change the mass balance, so the leaf
inflows $Q_{\mathrm{in}}$ are re-calibrated to hold the target outflow, and the
pressure field is recomputed
(Secs.~\ref{sec:valve},\ref{sec:leak},\ref{sec:node});\\
\textbf{(7)} detect and suppress nonphysical counterflow (lymph entering a terminal
afferent leaf) by a cycle-space block analysis, re-updating the feeding-leaf
inflows $Q_{\mathrm{in}}$, and where needed the local radii, so that every
afferent leaf carries net outflow into the network (Sec.~\ref{sec:bc});\\
\textbf{(8)} iterate the outer loop (5)--(7) as a self-consistent pressure
relaxation, repeatedly widening and re-anchoring the over-pressured edges
(driving the wall anchor pressure $P_{\mathrm{anchor}}\to0$ so the zero-tension
radius coincides with the relaxed radius) until the field stops dropping: every
edge radius outside the central thoracic duct and cisterna chyli settles within
$[0.05,1.50]$\,mm, $p_{\max}\le50$\,mmHg, and $Q_{\mathrm{out}}$ is preserved at
its calibrated value. The $95$th-percentile target ($\le30$\,mmHg) is not
enforced. The iteration stops once the pressure no longer decreases or touch the target.
Within this same outer loop, the feeding-leaf $Q_{\mathrm{in}}$ re-update also
rescues the few distal collector edges whose flow has fallen orders of magnitude
below their organ's bulk: their feeding leaf is raised until the edge reaches its
organ's typical range, with all leaves renormalised to hold $Q_{\mathrm{out}}$. A
small number ($<3\%$) of edges on anomalously long serial collector chains (peripheral) cannot be
recovered this way and are flagged as input-topology artefacts
(Fig.~\ref{fig:flow})
(Sec.~\ref{sec:res-relax}); \\
\textbf{(9)} validate the converged steady pressure field against independent
clinical data and quantify its sensitivity to the radius floor, the Murray
exponent $x_{\mathrm{Murray}}$, and the outlet impedance
(Secs.~\ref{sec:validation},\ref{sec:results}).

\subsection{Validation and sensitivity protocol}
\label{sec:validation}

Because the model is passive, the quantity we validate is the steady intraluminal
pressure distribution: $\Qout$ is an inflow-calibrated target
(Sec.~\ref{sec:res-massbal}), so its falling within $1.5$--$3.0$\,L/day \cite{GuytonHall2015,Moazzam2026} is by construction and reported only
for completeness, and the regional inflow magnitudes are prescribed inputs
(Eq.~(\ref{eq:qin}), \cite{YoffeyCourtice1970,Levick1991}), not observables. Pressures are
compared to the clinical cuff-occlusion ceilings (upper limb
$\le 39\pm14$\,mmHg \cite{Modi2007}; lower leg $\le 25.2\pm16.7$\,mmHg
\cite{Unno2010}) and to the thoracic-duct lumen pressure ($\sim$8\,mmHg
\cite{Lu2022,Moazzam2026}). Sensitivity is assessed by perturbing each Windkessel parameter
($R_p$, $R_d$, $C$) by $\pm20\%$ one at a time, and by sweeping the Murray
exponent $x_{\mathrm{Murray}}$ over $[2.4, 3.2]$.

\section{Results}
\label{sec:results}

\subsection{Radius hierarchy reproduces anatomy}
\label{sec:res-radius}

\begin{figure}[t]
  \centering
  \includegraphics[width=\linewidth]{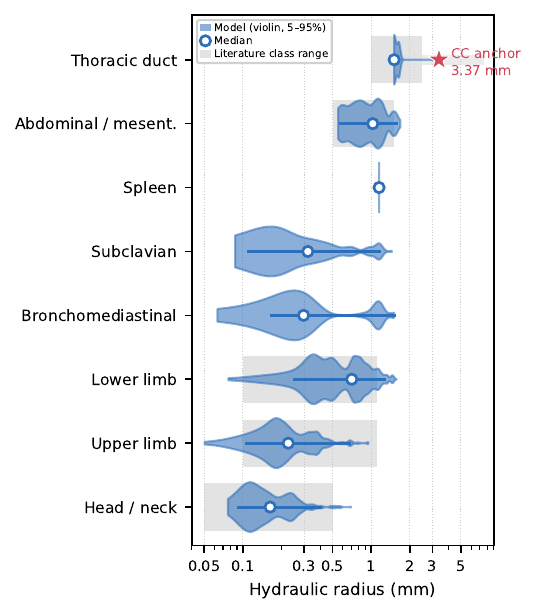}
  \caption{Distribution of post-cascade hydraulic radii by anatomical region
  (relaxed effective radius $r_{\mathrm{eff}}$ at the
  3\,L/day target), shown as per-region violins with the median (open circle)
  and 5--95th-percentile range, overlaid on the literature vessel-class band
  (gray). Regions are ordered central (top) to peripheral (bottom). The star
  marks the anchored cisterna-chyli peak (network maximum, 3.35\,mm). Percentiles
  reproduce Table~\ref{tab:radius}.}
  \label{fig:radius}
\end{figure}

The Murray cascade reproduces the expected anatomical calibre hierarchy across
all regions (Table~\ref{tab:radius}, Fig.~\ref{fig:radius}). The central path retains its anchored
values (thoracic duct median 1.50\,mm, cisterna chyli up to 3.35\,mm), the
abdominal/mesenteric and splenic trunks fall in the 0.5--1.5\,mm band, and the
peripheral limb and head/neck collectors are sub-millimetre at the median
(0.16--0.70\,mm), broadly consistent with the vessel-class reference ranges
\cite{Skandalakis2007,MargarisBlack2012,Pan2013}.

The limb radii fall within the
reported collector range (superficial lower-limb collectors span $0.2$--$2.2$\,mm
in diameter, i.e.\ radii to $\sim$1.1\,mm \cite{Pan2013}). The upper percentile of
the most heavily loaded lower-limb edges ($r_{95}\approx1.27$\,mm, $\sim$2.5\,mm
diameter) sits at the upper end of that range and marginally beyond it, reflecting
the self-consistent pressure relaxation widening those edges
(Sec.~\ref{sec:res-relax}). Across the $\sim$89{,}060 sub-edges the
hydraulic radius spans $[0.050, 3.35]$\,mm, from the global floor to the
anchored cisterna chyli. The number of branching levels from the trunk to the
leaves ($r_{\mathrm{leaf}}\approx r_{\mathrm{TD}}/2^{L/3}$) is consistent with the
$\sim$12--15 dichotomous generations implied by the floor.

\begin{table*}[t]
  \centering
  \caption{Post-cascade hydraulic radius by anatomical region (sub-edge
  percentiles), and the literature vessel-class reference range. The class range
  is expressed as luminal \emph{radius} (one-half the reported vessel diameter)
  for consistency with the model radii, with per-class sources: thoracic duct
  \cite{Skandalakis2007}; cisterna chyli \cite{Moazzam2022}; abdominal/mesenteric
  trunks \cite{Erden2005}; limb collectors \cite{Pan2013}; head/neck collectors
  \cite{MargarisBlack2012}. A range is listed only for classes with a published
  lymphatic caliber; it is omitted (--) for the spleen and the
  subclavian/bronchomediastinal terminal trunks, for which no quantitative vessel
  diameter is established. $n$ is the number of sub-edges.}
  \label{tab:radius}
  \begin{tabular}{lrrrrl}
    \toprule
    Region & $n$ & $r_{5}$ & $r_{50}$ & $r_{95}$ & Class range \\
           &     & (mm)    & (mm)     & (mm)     & (mm)        \\
    \midrule
    Thoracic duct (central) & 1145  & 1.50 & 1.50 & 1.51 & 1.0--2.5 (TD) \\
    Cisterna chyli (peak)   & --    & --   & --   & 3.35 & 2.5--7.5 (CC) \\
    Abdominal / mesenteric  & 4225  & 0.52 & 0.93 & 1.39 & 0.5--1.5 (trunk) \\
    Spleen                  & 90    & 1.05 & 1.05 & 1.05 & -- \\
    Subclavian              & 3855  & 0.11 & 0.32 & 1.15 & -- \\
    Bronchomediastinal      & 1940  & 0.16 & 0.29 & 1.50 & -- \\
    Lower limb              & 40615 & 0.23 & 0.65 & 1.20 & 0.1--1.1 (collector) \\
    Upper limb              & 29270 & 0.11 & 0.23 & 0.68 & 0.1--1.1 (collector) \\
    Head / neck             & 7835  & 0.09 & 0.16 & 0.40 & 0.05--0.5 \\
    \bottomrule
  \end{tabular}
\end{table*}

\subsection{Steady pressure field}
\label{sec:res-pressure}

The raw passive solve at the 3\,L/day delivery target is \emph{globally
over-pressured}. The median sub-edge transmural pressure is 45.9\,mmHg, with
75th, 95th, and 99th percentiles of 64.0, 123.8, and 517.2\,mmHg and a network
maximum of 1868\,mmHg. Nearly half of the sub-edges ($46.2\%$) exceed the 50\,mmHg clinical cuff scale
(Sec.~\ref{sec:validation}, Table~\ref{tab:validation}). The thoracic duct
itself carries a pressure gradient, from $8.1$\,mmHg at the venous-junction
end (the ostial-valve outlet, Sec.~\ref{sec:bc}) up to $\sim$38\,mmHg in the
abdominal cisterna chyli, and the body-wide afferent tree carries pressures
far above the upper-limb and lower-leg ceilings. 

Those over-pressure are
\emph{not} a localized distal radius-floor artifact of the kind seen in earlier,
lower-delivery calibrations of this model: it is a property of the whole afferent
network at this inflow scale, as three diagnostics show.
(i) No sub-edge sits on the radius floor at all (zero floor-pinned at
$r_{\rm floor}=0.05$\,mm), so the high pressures are carried by ordinary collector
generations, not by a thin floor-pinned tail.
(ii) The bulk pressure statistics are essentially \textit{insensitive} to the global
radius floor: sweeping $r_{\rm floor}=0.03$--$0.15$\,mm leaves the median at
$40.4$--$46.1$\,mmHg and the 95th percentile at $120$--$124$\,mmHg
(Table~\ref{tab:floor}), so no floor choice rescues the field from the
over-pressured regime. Only the network maximum responds strongly, while the net
outflow $\Qout$ holds at its calibrated value across the sweep.

(iii) Raising the distal afferent floors (head/neck, upper- and lower-limb) to an
anatomically appropriate $0.15$\,mm while keeping the $0.05$\,mm global floor
elsewhere (last row of Table~\ref{tab:floor}) lowers the network maximum
substantially (1868$\to$209\,mmHg) and
reduces the median (45.9$\to$40.4\,mmHg), but it does \emph{not} bring the field
into the physiological range: $40.9\%$ of sub-edges still exceed 50\,mmHg and the
95th percentile stays at 120\,mmHg. A single floor adjustment is therefore
inadequate at the 3\,L/day target. The over-pressure is a genuine network-level
feature of the unrelaxed static solve, which we resolve by the self-consistent
pressure relaxation of Sec.~\ref{sec:res-relax}.

\begin{table}[t]
  \centering
  \caption{Radius-floor sensitivity of the raw static solve (pre-relaxation) at
  the $3$\,L/day target (Sec.~\ref{sec:res-pressure}). Baseline is
  $r_{\rm floor}=0.05$\,mm; the last row is the region-aware floor defined there.}
  \label{tab:floor}
  \begin{tabular}{lrrrrr}
    \toprule
    Floor & $\Qout$ & $>\!50$ & $p_{\max}$ & $p_{50}$ & $p_{95}$ \\
    (mm)  & (L/day) &     (mmHg)         & (mmHg)     & (mmHg)   & (mmHg)   \\
    \midrule
    0.03                  & 3.00 & 46.3\% & 2034   & 46.1 & 124 \\
    \textbf{0.05} (base)  & 3.00 & 46.2\% & 1868   & 45.9 & 124 \\
    0.08                  & 3.01 & 46.2\% & 744    & 45.9 & 124 \\
    0.10                  & 3.01 & 44.8\% & 549    & 43.8 & 123 \\
    region $\sim 0.15$   & 3.01 & 40.9\% & 209    & 40.4 & 120 \\
    \bottomrule
  \end{tabular}
\end{table}

\subsection{Self-consistent pressure relaxation}
\label{sec:res-relax}

The globally over-pressured static field of Sec.~\ref{sec:res-pressure} is not
the operating state of the network: it is the fixed point of the passive solve
\emph{before} the over-narrow, over-pressured edges have been allowed to widen to
the load they carry. We therefore relax toward a self-consistent operating point.
The over-pressured edges are widened (Sec.~\ref{sec:perturbation}) and the wall
reference radius is re-anchored to the relaxed radius, so the passive wall tension
there vanishes ($P_{\mathrm{anchor}}\to0$), while the leaf inflows are re-updated
to preserve $Q_{\mathrm{out}}$. Each re-solve redistributes the load and can
drive a neighbouring edge above the pressure target, so the over-pressured set is
re-selected every pass and the iteration continues until the pressure field stops
decreasing.

The relaxation brings the network maximum down from the unphysically high value of
the raw static solve to 39.5\,mmHg, removes the over-50\,mmHg population entirely,
and compresses the whole distribution to a median
of 31.2\,mmHg with 75th, 95th, and 99th percentiles of 35.2, 37.5, and 38.9\,mmHg
(Table~\ref{tab:relax}). The per-region pressures then fall within the clinical
cuff scale (upper-limb 99th percentile 39.1\,mmHg, lower-leg 38.1\,mmHg).

The thoracic duct itself develops a physiological pressure gradient, from the
calibration-anchored junction value $\bar p_{\rm out}=7.9$\,mmHg at the
venous-junction outlet (see Sec.~\ref{sec:bc}) up to $\sim$27--29\,mmHg at the
abdominal cisterna chyli, consistent with the measured thoracic-duct pressure of
$8.2\pm3.2$\,mmHg \cite{Lu2022} (Sec.~\ref{sec:validation}).

We set $30$\,mmHg as a target for the relaxation drives toward, and it is placed below the clinical cuff-occlusion ceilings of
Sec.~\ref{sec:validation}, and the same level also selects which
edges are widened at each pass. The relaxation does not reach it. It ends after
$12$ outer passes with the maximum and the valve condition met, at $39.5$\,mmHg
and $\phi_{99}=9.37$, and with the $95$th percentile stalled at $37.5$\,mmHg.
Each later pass lowers it by less than the one before, so the field has reached
a floor set by the network rather than run out of iterations. We take it as the
pressure prediction of the static model. It lies below the upper-limb cuff
ceiling of $39\pm14$\,mmHg \cite{Modi2007} and above the lower-leg mean of
$25.2$\,mmHg while staying inside its $\pm16.7$ spread \cite{Unno2010}, so it
falls in the range at which collecting lymphatics are measured.

\begin{table}[t]
  \centering
  \caption{Transmural pressure distribution before and after the self-consistent
  relaxation (inflow calibrated to the $3$\,L/day target).}
  \label{tab:relax}
  \begin{tabular}{lrr}
    \toprule
    Quantity            & Pre-relax (static) & Post-relax (relaxed) \\
                        & (mmHg)             & (mmHg)                 \\
    \midrule
    Median $p_{50}$     & 45.9  & 31.2 \\
    $p_{75}$            & 64.0  & 35.2 \\
    $p_{95}$            & 123.8 & 37.5 \\
    $p_{99}$            & 517.2 & 38.9 \\
    Maximum $p_{\max}$  & 1868  & 39.5 \\
    Fraction $>\!50$\,mmHg & $46.2\%$ & $0\%$ \\
    \bottomrule
  \end{tabular}
\end{table}

\subsection{Per-vessel pressure and flow fields}
\label{sec:res-pqfields}

\begin{figure}[t]
  \centering
  \includegraphics[width=\linewidth]{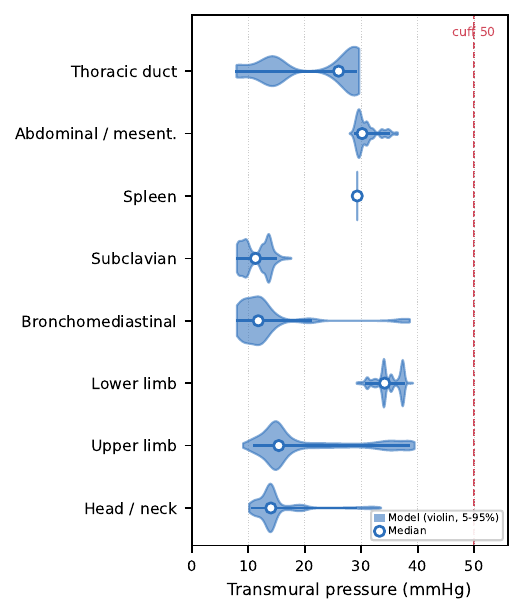}
  \caption{Per-region transmural pressure in the relaxed operating state
  (3\,L/day baseline), as per-region violins with the median (open circle) and
  5--95th-percentile range. The dashed line marks the $50$\,mmHg clinical cuff
  scale; the whole field lies below it. Regions are ordered central (top) to
  peripheral (bottom). The small high-pressure tails in the bronchomediastinal,
  upper-limb, and head/neck distributions arise from vessels far from the central
  outlet, where the resistive head accumulated along their drainage path raises the
  transmural pressure (see text).}
  \label{fig:pressure}
\end{figure}

\begin{figure}[t]
  \centering
  \includegraphics[width=\linewidth]{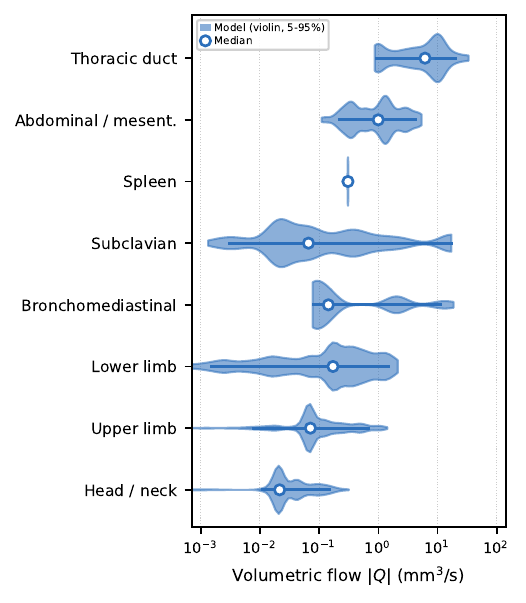}
  \caption{Per-region volumetric flow $|Q|$ (3\,L/day baseline, log scale), as
  per-region violins with median and 5--95th-percentile range, ordered central
  (top) to peripheral (bottom). Flow spans several decades, from the thoracic-duct
  outlet ($\Qout\approx34.8$\,mm$^3$/s) to the distal collectors; the low-$Q$
  left tails in the upper-limb and head/neck distributions are an input-topology
  artefact (anomalously long serial collector chains), not a physical effect (see
  text).}
  \label{fig:flow}
\end{figure}

Because the solve is vessel-resolved, the relaxed operating state
gives a transmural pressure, a volumetric flow, and a radius on
every sub-edge. Figures~\ref{fig:pressure} and
\ref{fig:flow} give the per-region distributions (the radius counterpart is
Fig.~\ref{fig:radius}). The pressure field is anatomically organised: the central thoracic duct
carries the lowest pressures and the peripheral lower-limb and abdominal collectors
the highest medians ($\sim$32--35\,mmHg), and every region lies below the $50$\,mmHg
clinical cuff scale (Fig.~\ref{fig:pressure}). This central-to-peripheral pressure ordering is a direct
consequence of the network geometry.

With the
afferent leaf inflows prescribed and the venous outlet fixed, the
transmural pressure at an edge is the accumulated viscous head built up along its
drainage path back from the outlet,
$P_e \approx p_{\rm out} + \sum_{e'\in\mathrm{path}(e)} Q_{e'} R_{e'}$. The distal (peripheral)
collectors sit at the upstream end of the longest chains of the narrowest segments,
so they accumulate the largest resistive head and therefore carry the highest
pressures, whereas the short, wide, outlet-adjacent central trunks accumulate the
least. The same mechanism produces the small high-pressure tails in Fig.~\ref{fig:pressure} seen within the
upper-limb, head/neck, and bronchomediastinal distributions: those sub-edges are the
most distal, narrowest links of unusually long serial collector paths, where the
$R\propto r^{-4}$ resistance accumulates over many generations. The elevated peripheral
pressures are thus a position-and-calibre (topology) effect, not a localized
artefact, and they remain bounded below the clinical cuff scale.

The volumetric flow, by contrast, is set by mass conservation, and it spans several orders of magnitude
(Fig.~\ref{fig:flow}, log scale). The thoracic duct carries the entire calibrated
outflow $\Qout\approx34.8$\,mm$^3$/s at its outlet. At each upstream bifurcation this
load is divided among the daughter branches, so the per-vessel flow falls by roughly
the branching ratio at every generation. The abdominal/mesenteric trunk and the
subclavian/bronchomediastinal collectors thus carry intermediate loads, while the
distal head/neck and limb collectors carry the smallest per-vessel flows. This
ordering mirrors the calibre hierarchy through the Murray closure $r\propto Q^{1/3}$
(Eq.~\ref{eq:branch}): the wide central vessels aggregate the flow of all their
descendants, whereas the narrow distal vessels each carry only a small share.

The few
sub-edges that fall orders of magnitude below their organ's bulk ($\approx$0.46\% of all sub-edges fall below $0.01$\,mm$^3$/s, excluding far peripheral vessels) are not a physiological feature but an input-topology artefact: a few drainage routes are encoded as implausibly long serial collector chains, e.g.~the head/neck path $v_{544}\!\to\!v_{566}\!\to\!\cdots\!\to\!v_{878}$ (11 serial segments, ${\approx}227$\,mm cumulative, where the node numbers are from Fig.~5 in \cite{Savinkov2020}), whose accumulated hydraulic resistance ($R\propto L$), compounded by transmural leakage along the path, deprives the terminal segment of flow. A more accurate human lymphatic-topology dataset, to be pursued in future work, would likely shorten these spurious chains and restore the flow of the affected edges. In the present graph only a handful of routes are involved, so they are negligible for the organ-level and whole-network results reported here.

\subsection{Whole-body mass balance: gross inflow to net output}
\label{sec:res-massbal}

With the inflow calibrated (Sec.~\ref{sec:validation}), the gross load is
$\sum_{\mathcal{R}} Q_{\mathcal{R}} = 65.75$\,mm$^3$/s ($5.68$\,L/day) and the
steady output $\Qout = 33.36$\,mm$^3$/s ($2.88$\,L/day) against the
$3.0$\,L/day calibration target. By
Eq.~(\ref{eq:massbalance}) the balance ($32.39$\,mm$^3$/s, $49.3\%$ of the gross
load) is reabsorbed transmurally (Eq.~(\ref{eq:leak}), $8.25$\,mm$^3$/s) and retained at
lymph nodes (Eq.~(\ref{eq:ln_retention}), $24.14$\,mm$^3$/s). 

The genuine mass-balance output is not this magnitude but the \emph{partition}.
Across a $2.0$--$4.0$\,L/day delivery sweep the achieved $\Qout$ tracks the target
almost exactly ($2.01$, $2.51$, $3.00$, $3.51$, $4.01$\,L/day) while the removed
fraction (nodal retention, dominant, plus transmural leak) stays nearly constant
at $46.8$--$47.6\%$ (Table~\ref{tab:lpload}), a scale-independent partition
coefficient as expected for a linear leak-and-retention network.
That sweep, the radius-floor sweep of Table~\ref{tab:floor}, the outlet
perturbations of Table~\ref{tab:sensitivity} and the
exponent sweep of Table~\ref{tab:murrayx} were computed on the baseline before
the Qin-outlier rescue of Sec.~\ref{sec:pipeline} was applied, so their rows are
internally consistent with one another and their baseline row sits at the
$3.00$\,L/day operating point rather than at the $2.88$\,L/day one reported
above. Its material
dependence is on the wall filtration coefficient $L_{\mathrm{leak}}$ and node retention: over a four-fold range
of $L_{\mathrm{leak}}$ the removed fraction rises monotonically
from $44.4\%$ to $51.7\%$ while $\Qout$ falls only from $3.16$ to $2.75$\,L/day,
the wall/valve nonlinearity partially buffering the change.

\begin{table}[t]
  \centering
  \caption{Two-axis sensitivity of the mass balance. Top block: net outflow
  targeted by a common multiplier on all $Q_{\mathcal{R}}$ ($L_{\mathrm{leak}}$ fixed). Lower
  block: $L_{\mathrm{leak}}$ varied at fixed inflow. ``Removed'' is the non-output fraction
  (transmural leak plus nodal retention).}
  \label{tab:lpload}
  \begin{tabular}{lrrr}
    \toprule
    Perturbation & Gross   & $\Qout$ & Removed \\
                 & (L/day) & (L/day) &         \\
    \midrule
    Target $2.0$\,L/day      & 3.77 & 2.01 & 46.8\% \\
    Target $2.5$\,L/day      & 4.72 & 2.51 & 46.9\% \\
    \textbf{Baseline $3.0$}  & 5.68 & 3.00 & 47.1\% \\
    Target $3.5$\,L/day      & 6.67 & 3.51 & 47.3\% \\
    Target $4.0$\,L/day      & 7.64 & 4.01 & 47.6\% \\
    \midrule
    $L_{\mathrm{leak}}\ 0.5\times$        & 5.68 & 3.16 & 44.4\% \\
    $L_{\mathrm{leak}}\ 1.0\times$ (base) & 5.68 & 3.00 & 47.1\% \\
    $L_{\mathrm{leak}}\ 2.0\times$        & 5.68 & 2.75 & 51.7\% \\
    \bottomrule
  \end{tabular}
\end{table}

\subsection{Sensitivity to the outlet boundary condition}
\label{sec:res-outlet}

We perturbed each three-element-Windkessel parameter ($R_p$, $R_d$, $C$) by
$\pm20\%$ about its baseline, one at a time (Table~\ref{tab:sensitivity}). The
net outflow is essentially invariant: $\Qout$ stays within $0.5\%$ of
3.00\,L/day across all perturbations, confirming that the whole-body outflow is
set by the internal network and the leak/retention
sinks.

The compliance column of Table~\ref{tab:sensitivity} is \emph{not} a physical
sensitivity. The exact steady outlet is the $C$-independent resistance
$g_{\mathrm{out}}=1/(R_p+R_d)$ (Eq.~(\ref{eq:rcr}), Sec.~\ref{sec:bc}). At the
adopted $C=10^{-15}$ the implicit-Euler $G_{\mathrm{eff}}$ is already pinned to this
direct-current limit, so a $\pm20\%$ change in $C$ leaves the converged solution
unchanged. The genuine outlet sensitivity is the $R_p$/$R_d$ response (i.e. of
$g_{\mathrm{out}}=1/R_{\mathrm{LVJ}}$). Unlike the superseded low-resistance
anatomical outlet, the ostial-valve resistance exerts real leverage on the
junction: a $\pm20\%$ change in $R_d$ (which is $\sim$77\% of $R_{\mathrm{LVJ}}$)
shifts the lymphovenous-junction pressure $\bar p_{\rm out}$ by $\pm0.76$\,mmHg
($7.34$--$8.86$\,mmHg about the $8.11$\,mmHg baseline), and a $\pm20\%$ change in
$R_p$ by $\pm0.22$\,mmHg. The bulk pressure field is far less affected (the
median shifts by $<0.5$\,mmHg and the 95th percentile by $<0.6$\,mmHg), because
the body-wide pressures (median $\sim$32\,mmHg after relaxation,
Sec.~\ref{sec:res-relax}) are set by the internal
network.

\begin{table}[t]
  \centering
  \caption{Outlet (ostial-valve RCR) sensitivity: each parameter perturbed
  $\pm20\%$, one at a time, on the $3.0$\,L/day baseline.}
  \label{tab:sensitivity}
  \begin{tabular}{llrrrr}
    \toprule
    Param. & Pert. & $\Qout$ & $\bar p_{\rm out}$ & $p_{50}$ & $p_{95}$ \\
           &       & (L/day) & (mmHg)             & (mmHg)   & (mmHg)   \\
    \midrule
    (baseline) & --      & 3.00 & 8.11 & 32.4 & 38.3 \\
    $R_p$ & $-20\%$       & 3.01 & 7.89 & 32.3 & 38.3 \\
    $R_p$ & $+20\%$       & 3.00 & 8.34 & 32.6 & 38.4 \\
    $R_d$ & $-20\%$       & 3.01 & 7.34 & 32.8 & 37.8 \\
    $R_d$ & $+20\%$       & 2.99 & 8.86 & 32.9 & 38.6 \\
    $C$   & $\pm20\%$     & 3.00 & 8.11 & 32.4 & 38.3 \\
    \bottomrule
  \end{tabular}
\end{table}

\subsection{Sensitivity to Murray's law}
\label{sec:res-murrayx}

In contrast to the outlet sweep, varying the Murray exponent over
$x_{\mathrm{Murray}}\in[2.4,3.2]$ produces a clean, monotone, physically interpretable trend
(Table~\ref{tab:murrayx}). At the production value $x_{\mathrm{Murray}}=3.0$ the solution
reproduces the baseline run to machine precision, an
implementation-consistency (regression) check on the exponent override, not a
physical result. The net outflow is almost insensitive to $x_{\mathrm{Murray}}$: it varies
only between 3.01\,L/day at $x_{\mathrm{Murray}}=2.4$ and 3.00\,L/day at $x_{\mathrm{Murray}}=3.2$ ($<1\%$),
holding at the calibrated delivery throughout. The exponent's dominant effect is
on the peripheral radius distribution and, through it, on the pre-relaxation
pressure field: a smaller $x_{\mathrm{Murray}}$ steepens the cascade attenuation (median sub-edge
radius 0.23\,mm at $x_{\mathrm{Murray}}=2.4$ versus 0.36\,mm at $x_{\mathrm{Murray}}=3.2$), which drives more leaf
tributaries onto the radius floor ($9.0\%$ of sub-edges with $r\le0.06$\,mm at
$x_{\mathrm{Murray}}=2.4$ versus $0.7\%$ at $x_{\mathrm{Murray}}=3.2$) and further inflates the already
over-pressured static field (median pressure 63.9\,mmHg and 99th percentile
1079\,mmHg at $x_{\mathrm{Murray}}=2.4$ versus 45.9 and 517\,mmHg at $x_{\mathrm{Murray}}=3.0$). The radius
floor $r_{\rm floor}$ and the exponent $x_{\mathrm{Murray}}$ thus jointly govern the magnitude of the
pre-relaxation over-pressure, while $\Qout$ is robust.  

It is \emph{not} evidence that $x_{\mathrm{Murray}}=3$ is the biologically correct exponent. The baseline value is motivated independently by the minimum-work and wall-shear arguments of Sec.~\ref{sec:murray}, which are cleanest in the large-vessel regime of the central trunk~\cite{Sherman1981}. What the sweep establishes is a direction. Reducing the exponent uniformly steepens the distal cascade, so the floor-pinned fraction and the pre-relaxation over-pressure both grow, and the sub-cubic exponents measured in blood vessels~\cite{Taylor2022Murray,Taylor2024MurrayMeta} cannot simply be adopted here as a global replacement. 

\begin{table}[t]
  \centering
  \caption{Murray-exponent single-parameter sweep (pre-relaxation static solve).
  $f_{\rm floor}$ is the fraction of sub-edges with $r\le0.06$\,mm.}
  \label{tab:murrayx}
  \begin{tabular}{rrrrrr}
    \toprule
    $x_{\mathrm{Murray}}$ & $\Qout$ & $r_{50}$ & $f_{\rm floor}$ & $p_{50}$ & $p_{99}$ \\
                          & (L/day) & (mm)     &                 & (mmHg)   & (mmHg)   \\
    \midrule
    2.4 & 2.97 & 0.226 & 9.0\% & 63.9 & 1079 \\
    2.6 & 2.99 & 0.259 & 5.4\% & 55.3 & 776 \\
    2.8 & 3.00 & 0.293 & 2.4\% & 49.9 & 597 \\
    3.0 & 3.00 & 0.327 & 1.4\% & 45.9 & 517 \\
    3.2 & 3.00 & 0.360 & 0.7\% & 41.6 & 415 \\
    \bottomrule
  \end{tabular}
\end{table}

\begin{table*}[t]
  \centering
  \caption{Comparison of the relaxed passive model (Sec.~\ref{sec:res-relax}) to
  independent physiological data. Clinical figures are inter-patient mean$\pm$SD,
  model entries single-realization percentiles, so agreement is qualitative (see
  text). The upper-limb and lower-leg clinical pressures are \emph{indirect}
  cuff-occlusion estimates (the external limb-cuff counter-pressure that gates
  fluorescent-tracer transit through the collector), i.e.\ occlusion ceilings
  rather than direct intraluminal measurements; only the thoracic-duct value is a
  lumen pressure.}
  \label{tab:validation}
  \begin{tabular}{llll}
    \toprule
    Quantity & Model & Clinical reference & Source \\
    \midrule
    Net outflow $\Qout$ & 2.88\,L/day & 1.5--3.0\,L/day (range) & \cite{GuytonHall2015,Moazzam2026}\\
    LVJ pressure $\bar p_{\rm out}$ (v996) & 7.9\,mmHg & $8.2\pm3.2$\,mmHg (porcine and human review study) & \cite{Lu2022,Ratnayake2018}\\
    Upper-limb $p$ (99th pct; max) & 39.1; 39.5\,mmHg & $\le 39\pm14$\,mmHg (ceiling) & \cite{Modi2007}\\
    Lower-leg $p$ (99th pct; max) & 38.1; 39.2\,mmHg & $\le 25.2\pm16.7$\,mmHg (ceiling) & \cite{Unno2010}\\
    \bottomrule
  \end{tabular}
\end{table*}

\section{Discussion}
\label{sec:discussion}

\subsection[Whole body model]{Whole body model}

We build the physical closures on the whole-body graph. The single anatomical edge is resolved into lymphangions and
sub-segments (Sec.~\ref{sec:anatomy}), its radius is fixed by a Murray law given
two independent derivations (Sec.~\ref{sec:murray}), and the constitutive parts
are an exponential passive wall law (Sec.~\ref{sec:wall}), hysteretic valves
(Sec.~\ref{sec:valve}), transmural leak (Sec.~\ref{sec:leak}), lymph-node
retention (Sec.~\ref{sec:node}) and a three-element venous outlet
(Sec.~\ref{sec:bc}). Rodent measurements are carried to human scale by the
allometric rescaling of Sec.~\ref{sec:species}. These enter one nodal Kirchhoff
system (Sec.~\ref{sec:hydraulic}), assembled in the fixed order of
Sec.~\ref{sec:pipeline} and closed self-consistently by the cycle-space radius
perturbation of Sec.~\ref{sec:perturbation}. The outcome is checked by the
protocol of Sec.~\ref{sec:validation}. What comes out is a pressure field and
not just a flow orientation as the predictions in our model, so it can be compared with clinical
measurement.

This work applies the reduced-order lymphatic literature rather than
competing with it. Those single-lymphangion and short-chain models supply the
constitutive physics assembled here, and we take that physics from the published
measurements. The passive wall
law is refit to the human pelvic pressure--diameter data of Athanasiou
\emph{et al.}~\cite{Athanasiou2017} (Sec.~\ref{sec:wall}), and the valve
thresholds are refit to the distension-dependent opening and closure pressures of
Davis \emph{et al.}~\cite{Davis2011}, with $R^2=0.95$ and $0.90$
(Fig.~\ref{fig: PvsX}). The sigmoid gating form and the closed-to-open resistance
ratio $\phi_{\mathrm{closed}}\approx370$ are inherited from the Bertram
series~\cite{Bertram2011,Bertram2014,Bertram2014a}, and the lymphangion spacing
from Margaris and Black~\cite{MargarisBlack2012}. What changes is the scale at
which these laws are evaluated. They were established on one vessel or on a few
in series~\cite{Razavi2017,jamalian2016network,ikhimwin2020computational,KimTithof2024}.
Here the same laws run on all $89{,}060$ sub-edges of an anatomical body-scale
network, where the topology sets each vessel's boundary condition.

\subsection[Drainage of the central nervous system]{Drainage of the central nervous system}

The model fixes a geometry, and gives a self-consistent set of vessel
radii for the normal physiological state, obtained from the Murray closure on a graph adapted under clinical-anatomist review.
The head and neck territory therefore drains through the cervical collectors and
the thoracic duct to the LVJ. It is the downstream conduit into
which cerebrospinal-fluid drainage terminates~\cite{du2024restoration}, and
which studies at the cervical scale currently have to prescribe as an outlet pressure.

The model gives a delivered fraction for the head
and neck route, and the same numbers show how thinly that territory is
represented in the graph. The superficial cervical inflow of
Table~\ref{tab:inflow} reaches the LVJ at a delivered
fraction of $30.0$\%, against $53$ to $76$\% for the visceral territories, so
most of the head and neck load is taken up by leak and nodal retention on the
way down and the territory supplies $1.3$\% of the outlet flow. The deep
cervical compartment, which receives meningeal drainage, carries
$0.023$\,mm$^3$/s of the $65.75$\,mm$^3$/s gross inflow, so the part of the
network closest to the central nervous system is also its least resolved
part, and a cervical drainage prediction
depends on the graph that human anatomical data does not yet fix.

An independent model of the same territory gives a first external check.
Kim and Tithof simulate cerebrospinal-fluid efflux from the murine skull through
the cervical lymphatic vessels with a contraction-driven lumped
model~\cite{KimTithof2024}, built for a different purpose and a different
species. Their validation window is a per-vessel flow of $0.012$ to
$0.034$\,$\mu$L/min at a median diameter of $76$ to $88$\,$\mu$m. Carried to
human scale by the mouse factors of Eq.~\eqref{Eq: rescale}, a Poiseuille flow
ratio $\alpha_p^m(\lambda_r^m)^4/\lambda_L^m=112$ and a diameter ratio
$\lambda_r^m=4.48$, that becomes $0.022$ to $0.063$\,mm$^3$/s per vessel at a
radius of $0.170$ to $0.197$\,mm. The head and neck collectors here carry a
median of $0.022$\,mm$^3$/s at a median radius of $0.162$\,mm. The flow ratio
depends on the fourth power of $\lambda_r^m$, and its one-sigma corners span $4.5$
to $1102$, so the flow agreement is only a consistency check. The radius
comparison does not carry that amplification.

The same comparison shows what a whole-body network adds to a
cervical-scale one. \cite{KimTithof2024} terminate their modelled vessel with a venous blood pressure prescribed at a
truncated boundary for murine at the superficial cervical
lymph node, $10$\,mm from the nasal mucosa, and impose
a central venous outlet pressure of $6$\,mmHg there, swept to $10.44$\,mmHg in
their sensitivity analysis, so the lymphatic path from that node
to the LVJ is not represented. In our
model that path is resolved and used human scenario. Head and neck lymph reaches the subclavian trunk
through mainly four edges, and they are distal vertex lies
outside the head and neck territory. Those distal vertices are v906, v910, v477
and v474 (see Fig.~5 in \cite{Savinkov2020}), at luminal pressures of $10.2$, $13.6$, $15.5$ and $17.6$\,mmHg, and
the edges carry $0.317$, $0.280$, $0.093$ and $0.092$\,mm$^3$/s, respectively. The two dominant
collectors are the edges ending at v906 and v910, whose $0.597$\,mm$^3$/s is
$76$\% of the $0.783$\,mm$^3$/s leaving the territory. Pressure falls from
there to $7.9$\,mmHg at the LVJ
(Table~\ref{tab:validation}).

The model can be asked where a change in central venous pressure is
felt. After traumatic brain injury an adrenergic surge raises that pressure and
lowers cervical lymphatic contractility together, and blocking the surge
normalises both and reduces cerebral oedema \cite{Hussain2023}. Central venous
pressure is an explicit boundary of this model, $P_{\mathrm{subclav}}$ of
Sec.~\ref{sec:bc}, so changing it is an input here. We study this topic with another research paper, and the reason is a modelling requirement. The response to a raised venous
pressure is set by three mechanisms acting together: the lymphangion pump, the
transmural leak that returns fluid to the tissue, and a lymphovenous ostial valve
that opens and closes with the breath. The present model holds the
first at zero by construction, carries the second, and replaces the third with a
steady three-element outlet. Reporting a propagation distance from it would therefore
describe a network that can neither pump against the raised load
nor gate its own outlet. The companion study adds the contraction
machinery and the respiration-gated ostial valve, and it answers this question
on the geometry, the calibre hierarchy and the passive pressure field
established here.

\subsection[Addtional contributions]{Addtional contributions}

The value of these predictions is that they are reached
on a human-scale anatomical graph with literature-grounded constitutive
physics. The removed fraction ($\sim$49\%, nearly independent of the outflow
scale) counts as evidence in the same way, while the inflow-calibrated outflow is a prescribed
target. The sensitivity analysis shows the bulk field to be governed by the
internal network and insensitive to the outlet model, and the raw over-pressure is
a network-level property, resolved by the self-consistent relaxation (which lands
slightly above the $30$\,mmHg collector target), not a distal-floor artifact.

Additional contribution is that two independent quantities land
in clinical range at the same operating point without either being fitted, the
radius from the Murray closure and the transmural pressure from the nodal solve
(Tables~\ref{tab:radius} and \ref{tab:relax}). Two of those values sit at the
edge of their published ranges (Table~\ref{tab:validation}).

Because the framework is whole-body and modular, it is a natural substrate for
clinically oriented extensions. Adding tracer transport would yield regional
time--activity curves of the kind measured by lymphoscintigraphy or ICG/NIRF
imaging, and switching on the active-contraction machinery would let the same
network represent disease states, such as the pressure asymmetry, stasis, and
valvular counterflow associated with obstruction and lymphedema, both vessel by
vessel and after aggregation to clinically observed regional readouts.

\subsection[Limitations]{Limitations}

The first limitation is that the Murray exponent is held at a single
value. The cube law is a theoretical
optimum obtained for laminar flow \cite{Sherman1981}, and admitting non-ideal
flow already lowers it toward $7/3$ \cite{Uylings1977}. Recent human coronary
estimates sit below the cubic value, at $2.15$ to $2.38$ in epicardial arteries
\cite{Taylor2022Murray} and at a pooled $2.39$ with a $95$\% interval of $2.24$
to $2.54$ \cite{Taylor2024MurrayMeta}, although the reported vascular range is
wide and individual morphometric studies reach and exceed $3$. The
lymphatic-specific evidence points lower still at the distal end. Murine dermal
lymphatic capillaries give $1.44\pm0.06$, against $2.21\pm0.10$ for the blood
capillaries of the same tissue \cite{Talkington2022Lymphatic}. No exponent is
established by direct measurement for human collecting lymphatics or for the
trunks that carry most of the load here (Sec.~\ref{sec:murray}).

A calibre-dependent exponent, nearer the cube law on the central trunks
and lower on the distal collectors, is therefore the physiologically motivated
alternative to the constant we use. It is not what the sweep of
Table~\ref{tab:murrayx} tests, since that sweep moves the exponent uniformly
over the whole network. What the sweep does show is how the distal end responds
to a lower exponent: reducing $x_{\mathrm{Murray}}$ from $3.0$ to $2.4$ drops the
median sub-edge radius from $0.327$ to $0.226$\,mm, drives the floor-pinned
fraction from $1.4$\% to $9.0$\%, and raises the pre-relaxation median pressure
from $45.9$ to $63.9$\,mmHg. A genuinely lower distal exponent would therefore
push the peripheral collectors further onto the radius floor, so the floor and
the exponent cannot be chosen independently. Resolving that coupling needs an
exponent measured on human collecting lymphatics, which does not yet exist.

A model of cervical lymphatic drainage adopts the same low exponent for
its initial vessels and finds it functionally consequential~\cite{KimTithof2024}.
Under raised intracranial pressure the cube law allows the vessel diameter to swing
between about $110$ and $330$\,$\mu$m, while at $1.45$ it stays between about
$100$ and $170$\,$\mu$m. That exponent is taken from the same murine measurement
cited above, so it is not an independent value, though it is still evidence that
the choice changes what a model predicts.

The second limitation is that the lymphangions do not pump. Active circumferential
tension is switched off throughout, by construction (Sec.~\ref{sec:anatomy}),
so the model describes the passive hydraulic skeleton and not the
contraction-driven transport that the vessels perform in life. The
contraction-driven companion study is where that is addressed.

The third limitation is posture and the external tissue pressure. We set
$P_{\mathrm{ext}}=0$, so the
transmural and intraluminal pressures coincide (Sec.~\ref{sec:leak}), and the
organ-level loading that posture imposes is absent. That loading is
patient-specific and there is not enough data to prescribe it in general.
A lymphedematous state is nevertheless reachable within the present static
model by raising $P_{\mathrm{ext}}$ on a chosen territory or by clearing its
nodes, which is a further use of the boundary conditions the model already
carries.

Extrinsic pumping is not switched off in the same way.
A respiration-driven pressure difference of roughly $\sim 15$\,cmH$_2$O, about $11$\,mmHg,
is imposed along the cisterna-chyli trunk (Sec.~\ref{sec:bc}), so the model
carries the respiratory pump while the intrinsic contraction is absent. That
difference is of the same order as the thoracic-duct pressure rise reported in
Sec.~\ref{sec:res-relax}, so the central return here is driven in part by an
imposed boundary term and not by the network alone. Separating the two
contributions requires a run with the respiratory term removed, which is not
reported here.

Four narrower limitations remain. The inter-valve spacing is a single
global $3$\,mm while the one direct measurement on a large trunk gives about
$31$\,mm, an order of magnitude apart. The rodent-to-human transfer of
$L_{\mathrm{leak}}$, $k_1$, $k_2$ and $R_{\mathrm{node}}$
(Sec.~\ref{sec:species}) is allometric. The sub-segment
count $N_g=5$ makes the discrete mass balance exactly mesh-independent
(Sec.~\ref{sec:anatomy}), which does not by itself establish that the pressure
field has converged, and no mesh-convergence study is reported here. Finally
the relaxed field lands slightly above the $30$\,mmHg collector target
(Sec.~\ref{sec:res-relax}), which is the one target the model does not meet.

\subsection[Outlook]{Outlook}

We frame the present result as physiological \emph{consistency}, and sharpening it is primarily a measurement problem. The single global Murray
exponent should give way to a measured, spatially varying one. The sweep of
Table~\ref{tab:murrayx} shows only that the exponent controls the peripheral
radius and the pre-relaxation over-pressure, and that a uniform reduction below
$3$ makes both worse, so the spatial profile has to come
from data. Region-specific human wall-law parameters
($k_1,k_2$), nodal resistances, and the filtration coefficient $L_{\mathrm{leak}}$
would replace the present allometric rescaling from rodents. Direct human data
on axial wall stretch, valve spacing, and external tissue pressure
$P_{\mathrm{ext}}$ would remove the model-anchored wall reference state. Each is an
experimentally accessible target whose measurement would convert a consistency
check into a quantitative validation.

\section{Conclusions}
\label{sec:conclusion}

We have presented a whole-body, anatomically explicit passive hydraulic model of
the human lymphatic network. Its central result is a complete vessel-resolved
hydraulic map: the steady transmural pressure, volumetric flow, and luminal radius
on every one of the $\sim$89{,}060 vessel segments of an anatomical body-scale
graph. Vessel radii are closed by a Murray cube law with a dual physical
derivation, and the pressure and flow fields are solved by nodal
hydraulic balance with experimentally informed wall mechanics, valves, transmural
leakage, lymph-node retention, and a three-element venous outlet.

The impact of this work is that it moves the unit of lymphatic modelling
from the vessel to the body. Wiring the anatomical territories together and
closing them with a single nodal Kirchhoff system means the pressure and flow on
every one of the 89{,}060 segments follow from the whole
network. Only the inflow scale is calibrated, and it is
set so that the net outflow is physiological. The pressures and the radii are not fitted to the data they
are later compared against. The pressures stay below the clinical cuff
ceilings, and the radii land inside the published vessel-class range in every
region that has one. Calibre and pressure reach their ranges by different
routes, the Murray closure and the nodal solve, so they were free to disagree
with the anatomy and with each other. About $49$\% of the gross lymph load
is removed before the outlet, and a delivery sweep shows that fraction to
be nearly independent of the outflow scale. The construction is modular, so it
is a base for finer whole-body simulations.

The clinical value of a whole-body baseline is that it makes
patient-specific questions computable. The graph is anatomical, so the regional
priors of Table~\ref{tab:inflow} can in principle be replaced by an individual
patient's imaging, and the solve then returns pressure and flow vessel by
vessel.

Two classes of question follow from boundary conditions the model already
carries. The Fontan circulation and the post-traumatic state share a raised
central venous pressure, which enters the model through the outlet far field whose
sensitivity is quantified in Sec.~\ref{sec:res-outlet}. Removing a group of
lymph nodes, as in an axillary dissection, is instead a change to the graph, and
a whole-body solve returns the redistribution of flow that follows. Neither scenario is attempted here. Both are changes of
input: the outlet far field is a single number, and a
nodal clearance is an edge set. The solve reported above is what turns either of
them into a body-wide pressure and flow field.

\section*{List of abbreviations}
CC: cisterna chyli;
CNS: central nervous system;
CSF: cerebrospinal fluid;
GI: gastrointestinal;
ICG: indocyanine green;
LN: lymph node;
LVJ: lymphovenous junction;
NIRF: near-infrared fluorescence;
RCR: resistance--compliance--resistance (three-element Windkessel);
TD: thoracic duct.

\section*{Declarations}

\subsection*{Ethics approval and consent to participate}
Not applicable.

\subsection*{Consent for publication}
Not applicable.

\subsection*{Availability of data and materials}
Data and materials are available from the authors following any reasonable request.

\subsection*{Competing interests}
The authors declare no competing interests.

\subsection*{Funding}

\subsection*{Authors' contributions}
Conceptualization: CFC; Methodology: CFC; Formal analysis and investigation:
CFC, BA, ES; Writing - original draft preparation: CFC; Writing - review and editing:
CFC, BA, ES, SM; Funding acquisition: SM; Resources: SM; Supervision: SM.

\subsection*{Acknowledgements}
Not applicable.

\bibliographystyle{vancouver}
\bibliography{References}

\end{document}